\documentclass[largeformat]{interact}

\usepackage{hyperref}         % for hypertext (\href, \url, BibTeX url=, etc.)

\usepackage[utf8]{inputenc}                   % input encoding (from template)

\usepackage[numbers,sort&compress]{natbib}   % for bib support (from template)

\newcommand{\ehat}{\bm e}

\newcommand{\ex}{\ehat_x}
\newcommand{\ey}{\ehat_y}
\newcommand{\ez}{\ehat_z}

\newcommand{\nhat}{\bm n}

\newcommand{\neps}{\nhat_\epsilon}
\newcommand{\phieps}{\varphi_\epsilon}

\newcommand{\nuhat}{\bm \nu}

\newcommand{\bmp}{\bm p}
\newcommand{\bmu}{\bm u}
\newcommand{\bmv}{\bm v}

\newcommand{\bmB}{\bm B}
\newcommand{\bmD}{\bm D}
\newcommand{\bmE}{\bm E}
\newcommand{\bmF}{\bm F}
\newcommand{\bmH}{\bm H}
\newcommand{\bmM}{\bm M}
\newcommand{\bmP}{\bm P}

\newcommand{\bmzero}{\bm 0}

\newcommand{\bfI}{\mathbf{I}}
\newcommand{\bfP}{\mathbf{P}}

\newcommand{\chipara}{\chi_{\scriptscriptstyle\parallel}}
\newcommand{\chiperp}{\chi_{\scriptscriptstyle\perp}}

\newcommand{\chia}{\chi_\text{a}}

\newcommand{\chie}{\chi^\text{e}}

\newcommand{\chiepara}{\chie_{\scriptscriptstyle\parallel}}
\newcommand{\chieperp}{\chie_{\scriptscriptstyle\perp}}

\newcommand{\chitensor}{\bm\chi}

\newcommand{\eps}{\varepsilon}

\newcommand{\epara}{\eps_{\scriptscriptstyle\parallel}}
\newcommand{\eperp}{\eps_{\scriptscriptstyle\perp}}

\newcommand{\epsa}{\eps_\text{a}}

\newcommand{\epstensor}{\bm\eps}

\newcommand{\phim}{\varphi_\text{m}}

\newcommand{\calF}{\mathcal{F}}

\newcommand{\calFE}{\calF_E}
\newcommand{\calFH}{\calF_H}

\newcommand{\Ec}{E_\text{c}}
\newcommand{\Hc}{H_\text{c}}
\newcommand{\Vc}{V_\text{c}}

\newcommand{\We}{W_\text{e}}

\newcommand{\sigmaf}{\sigma_\text{f}}

\newcommand{\thz}{\theta_{\!,z}}

\newcommand{\varphiz}{\varphi_{\!,z}}

\newcommand{\wz}{w_{\!,z}}

\newcommand{\rhoP}{\rho_P}
\newcommand{\sigmaP}{\sigma_P}

\newcommand{\cth}{\cos\theta}
\newcommand{\ccth}{\cos^2\!\theta}
\newcommand{\sth}{\sin\theta}
\newcommand{\ssth}{\sin^2\!\theta}

\newcommand{\Fbar}{\bar{\calF}}

\newcommand{\zbar}{\bar{z}}

\newcommand{\thbar}{\bar{\theta}}

\newcommand{\ssthbar}{\sin^2\!\thbar}

\newcommand{\aK}{\alpha_K}
\newcommand{\aeps}{\alpha_{\eps}}

\newcommand*{\Vmax}{V_{\max}}
\newcommand*{\Vmin}{V_{\min}}
\newcommand*{\Vth}{V_{\text{th}}}
\newcommand*{\VLP}{V_{\min}^{\text{theor}}}
\newcommand*{\VUP}{V_{\text{th}}^{\text{theor}}}
\newcommand*{\VBP}{V_{\max}^{\text{theor}}}

\newcommand{\td}{\text{d}}

\newcommand{\Freed}{Fr\'{e}edericksz}

\let\div\relax                                      % must undefine \div first
\DeclareMathOperator{\div}{div}

\DeclareMathOperator{\curl}{curl}

\DeclareMathOperator{\tr}{tr}

\begin{document}

%\articletype{ANONYMOUS VERSION}
%\articletype{LONG VERSION}

\title{Anomalous behavior of electric-field \Freed\ transitions}

\author{\name{Eugene C. Gartland, Jr.}
  \affil{Department of Mathematical Sciences, Kent State University,
    Kent, OH 44242, USA}}

\thanks{CONTACT E. C. Gartland, Jr.,
  \href{mailto:gartland@math.kent.edu}{\nolinkurl{gartland@math.kent.edu}},
  \url{http://www.math.kent.edu/\string~gartland}}

\maketitle

\begin{abstract}
  \Freed\ transitions in nematic liquid crystals are re-examined with
  a focus on differences between systems with magnetic fields and
  those with electric fields.  A magnetic field can be treated as
  uniform in a liquid-crystal medium; while a nonuniform director
  field will in general cause nonuniformity of the local electric
  field as well.  Despite these differences, the widely held view is
  that the formula for the threshold of local instability in an
  electric-field \Freed\ transition can be obtained from that for the
  magnetic-field transition in the same geometry by simply replacing
  the magnetic parameters by their electric counterparts.  However, it
  was shown in [Arakelyan, Karayan, and Chilingaryan, \textit{Sov.\
    Phys.\ Dokl.}, \textbf{29} (1984) 202--204] that in two of the six
  classical electric-field \Freed\ transitions, the local-instability
  threshold should be \emph{strictly greater} than that predicted by
  this magnetic-field analogy.  Why this elevation of the threshold
  occurs is carefully examined, and a simple test to determine when it
  can happen is given.  This ``anomalous behavior'' is not restricted
  to classical \Freed\ transitions and is shown to be present in
  certain layered systems (planar cholesterics, smectic~A) and in
  certain nematic systems that exhibit periodic instabilities.
\end{abstract}


%%%%%%%%%% BODY %%%%%%%%%%

\section{Introduction}

\Freed\ transitions are the most basic instability in the study of
liquid crystals.  In simplest terms, a \Freed\ transition occurs when
an externally applied magnetic or electric field reaches a strength
that is sufficient to distort the uniform orientational equilibrium
state of a liquid-crystal sample, the ground-state configuration
having been imposed by anchoring conditions on confining substrates.
Different orientations of the liquid-crystal ground state and the
external field give rise to different types of \Freed\ transitions.
This phenomenon has been studied for decades in a large number of
variations and is discussed in all the standard texts---see
\cite{pikin:91,chandrasekhar:92,degennes:prost:93,virga:94,stewart:04},
where references to the historical literature can also be found.  Here
we focus on differences that can occur between magnetic-field and
electric-field \Freed\ transitions with respect to \emph{local
  instability thresholds}, differences that are little known and not
well understood.

We consider the simplest liquid-crystal phase, an achiral uniaxial
nematic, modeled at the macroscopic level in terms of the continuum
theory of Oseen, Zocher, and Frank.  The free energy of the system is
expressed in terms of an integral functional of the nematic director
$\nhat$, a unit-length vector field representing the average
orientation of the distinguished axis of the anisometric molecules in
a fluid element at a point.  Intermolecular forces encourage local
parallel alignment of $\nhat$.  In the uniaxial nematic phase, the
material is \emph{transversely isotropic}, the magnetic and electric
susceptibilities having one value for fields aligned parallel to
$\nhat$ and another value for fields perpendicular to it.  If the
anisotropy is \emph{positive} (the susceptibility parallel to $\nhat$
greater than that perpendicular to it), then the external field will
exert a couple on $\nhat$ encouraging it to align \emph{parallel} to
the external field; while if the anisotropy is \emph{negative}, the
field will encourage the director to orient \emph{perpendicular} to
the field.

Magnetic fields are influenced by the presence of a liquid-crystal
medium.  For the typical parameter values of such materials, however,
this influence is negligible.  Thus a magnetic field in a liquid
crystal can be treated as a \emph{uniform} external field.  An
electric field is influenced by a liquid-crystal medium as well but
with a much stronger coupling---we make this more precise in the next
section.  The distinction between these two cases is discussed in
\cite[Sec.\,IV.1]{pikin:91}, \cite[Sec.\,3.4.1]{chandrasekhar:92},
\cite[Sec.\,3.3]{degennes:prost:93}, \cite[Sec.\,4.1]{virga:94},
\cite[Sec.\,2.3]{stewart:04}, and
\cite{arakelyan:karayan:chilingaryan:84,gartland:21}.  Thus the
equilibrium state of a liquid crystal subject to an electric field
should be determined in a self-consistent way, with the director field
and the electric field treated as \emph{coupled} state variables.  In
general this coupling leads to \emph{inhomogeneity} of the electric
field and complicates the determination of equilibrium fields and the
assessment of their local stability properties.

While the differences between magnetic fields and electric fields in
liquid crystals have been appreciated for some time, the widely held
view is that they only give rise to modest \emph{quantitative}
differences but not to \emph{qualitative} differences in the context
of instabilities such as \Freed\ transitions.  For example, in
\cite[Sec.\,IV.1]{pikin:91} (referencing
\cite{gruler:meier:72,gruler:scheffer:meier:72}),
\cite[Sec.\,3.3.1]{degennes:prost:93} (referencing
\cite{gruler:meier:72}), and \cite[Sec.\,3.5]{stewart:04}, it is
asserted that electric-field \Freed\ thresholds can be obtained from
the formulas for magnetic-field thresholds simply by substituting the
electric parameters for the corresponding magnetic parameters.  In
fact, this was borne out in \cite{gruler:scheffer:meier:72} (recounted
in \cite[Sec.\,IV.1]{pikin:91}) and in \cite{deuling:72} (recounted in
\cite[Sec.\,3.5]{stewart:04}) for the case of the electric-field
splay-\Freed\ transition, and this has contributed to the almost
universal acceptance of this ``magnetic-field analogy.''

Contrary to the above, in \cite{arakelyan:karayan:chilingaryan:84} (a
paper that is not very well known) it was shown that both the
electric-field bend-\Freed\ transition with a positive dielectric
anisotropy and the splay transition with a negative anisotropy have
instability thresholds that are \emph{strictly greater} than those
predicted by simply replacing the magnetic parameters by their
electric counterparts.  We confirmed this result in \cite{gartland:21}
and further showed that this effect is more general and is also ``one
sided'': taking into account the coupling between the electric field
and the director field can \emph{elevate} an electric-field-induced
instability threshold or leave the threshold \emph{unchanged}, but it
can \emph{never lower} it.  In \cite{gartland:21} we also developed a
simple criterion (based on problem geometry and field orientations) to
identify situations in which such a threshold-elevating effect would
take place.  Details are provided in what follows.

The two \Freed\ transitions mentioned above (the bend transition with
a positive anisotropy and the splay transition with a negative
anisotropy) are closely related in their modeling and are the only
ones of the six classical electric-field transitions that exhibit this
anomalous behavior---the other four transitions follow the
magnetic-field analogy.  The purpose of the present paper is to
explain this and to illustrate how it fits into the framework of a
more general theory of electric-field-induced instabilities (as
developed in \cite{gartland:21}).  Circumstances that enable the
electric-field transition to be \emph{first order} emerge from the
analysis as well.  This work was motivated by results of experiments
reported in \cite{frisken:palffy:89,frisken:palffy:89b} and related
numerical investigations presented in \cite{richards:06}.  In addition
to these \emph{classical} \Freed\ transitions, certain ``generalized
\Freed\ transitions'' exhibit this anomalous behavior---these include
periodic instabilities in certain nematic systems and in some layered
systems (planar cholesteric and smectic-A films)---and these are
discussed as well.

The paper is organized as follows.  In
Section~\ref{sec:free-energies}, free energies that can be used to
model the magnetic-field and electric-field \Freed\ transitions are
given, and the differences between them are discussed.  The classical
magnetic-field transitions are reviewed in
Section~\ref{sec:H-field-transitions}, with the local stability of
equilibrium states studied from a variational point of view.  The
modeling of the analogous electric-field transitions is considered in
Section~\ref{sec:models-of-E-field-transitions}, pointing out the
differences among the cases that result from the different geometries.
In Section~\ref{sec:analysis-of-E-field-bend}, the electric-field
bend-\Freed\ transition with positive dielectric anisotropy is
carefully analyzed, and the formula for the elevated local-instability
threshold predicted in \cite{arakelyan:karayan:chilingaryan:84} is
re-derived using a perturbation expansion.  A necessary condition for
the local stability of an equilibrium director field and coupled
electric field in a general setting is given in
Section~\ref{sec:E-field-induced-instabilities}.  This condition
provides a ``litmus test'' for when an instability threshold will be
altered and also illustrates why a threshold can only be elevated, not
lowered.  Applications to both classical and generalized \Freed\
transitions illustrating the utility of this test are discussed.  In
Section~\ref{sec:comparison-with-experiment}, results of experiments
reported in \cite{frisken:palffy:89} are shown to be in qualitative
agreement with predictions here.  The main findings are summarized in
the concluding Section~\ref{sec:conclusions}.

\section{Free energies: electric fields versus magnetic fields}

\label{sec:free-energies}

Equilibrium states of the systems we study are characterized as
stationary points of appropriate free-energy functionals.  For
magnetic fields generated by current-stabilized electromagnets or
electric fields produced by electrodes held at constant potential, the
free energies have the general forms
\begin{equation*}
  \calFH = \int_\Omega \Bigl( \We - \frac12 \bmB\cdot\bmH \Bigr) \td V , \quad
  \calFE = \int_\Omega \Bigl( \We - \frac12 \bmD\cdot\bmE \Bigr) \td V ,
\end{equation*}
where $\Omega$ is the region occupied by the liquid crystal.  Here
$\bmB$ is the magnetic induction, $\bmH$ the magnetic field, $\bmD$
the electric displacement, $\bmE$ the electric field, and $\We$
denotes the distortional elasticity, which can be written in the form
of the classical Frank formula
\begin{equation}\label{eqn:We}
  \We =
  \frac{\,K_1}2 (\div\nhat)^2 + \frac{\,K_2}2 (\nhat\cdot\curl\nhat)^2 +
  \frac{\,K_3}2 |\nhat\times\curl\nhat|^2 +
  K_{24} \bigl[ \tr(\nabla\nhat)^2 - (\div\nhat)^2 \bigr] .
\end{equation}
See \cite[Sec.\,3.2]{virga:94} or \cite[Sec.\,2.2]{stewart:04}.  The
elastic constants $K_1$, $K_2$, $K_3$, and $K_{24}$ are material and
temperature dependent.  For a transversely isotropic medium in the
linear regime, the magnetic quantities can be taken in the form
\begin{equation*}
  \bmB = \mu_0 ( \bmM + \bmH ) , \quad
  \bmM = \chitensor(\nhat) \bmH , \quad
  \chitensor = \chiperp \bfI + \Delta \chi ( \nhat\otimes\nhat ) , \quad
  \Delta \chi := \chipara - \chiperp ,
\end{equation*}
where $\mu_0$ is the free-space permeability, $\bmM$ the
magnetization, $\chitensor$ the susceptibility tensor, and $\chipara$
and $\chiperp$ the susceptibilities parallel to $\nhat$ and
perpendicular to $\nhat$ (with values in the range
$|\chiperp|, |\chipara| \approx 10^{-6}\text{--}10^{-5}$ in SI units
for liquid-crystal materials, \cite[Sec.\,3.2.1]{degennes:prost:93}).
Using these expressions, the equations of magnetostatics in the
absence of any current density ($\curl\bmH=\bmzero$, $\div\bmB=0$)
guarantee the existence of a magnetic scalar potential $\phim$ that
must satisfy
\begin{equation*}%\label{eqn:divBzero}
  \Delta \phim + \frac{\Delta\chi}{1+\chiperp}
  \div \bigl[ \bigl(\nabla\phim \cdot \nhat \bigr) \nhat \bigr] = 0 , \quad
  \bmH = - \nabla \phim .
\end{equation*}
For typical liquid crystals, $\Delta\chi/(1+\chiperp)\approx10^{-6}$
(see \cite[Sec.\,3.2.1]{degennes:prost:93}), and the term involving
$\nhat$ above is negligible---the magnetic field is uninfluenced by
the director field to a high degree of approximation.

The electric quantities for a medium such as ours have forms that are
similar to those for the magnetic quantities, though they are usually
expressed in slightly different notation:
\begin{equation}\label{eqn:eps-tensor}
  \bmD = \epstensor(\nhat) \bmE , \quad
  \epstensor = \eps_0
  \bigl[ \eperp \bfI + \epsa (\nhat\otimes\nhat ) \bigr] , \quad
  \epsa := \epara - \eperp ,
\end{equation}
where $\eps_0$ is the free-space permittivity, $\epstensor$ the
dielectric tensor, and $\epara$ and $\eperp$ the permittivities
parallel to $\nhat$ and perpendicular to $\nhat$.  Assuming there is
no distribution of free charge, the equations of electrostatics
($\curl\bmE=\bmzero$, $\div\bmD=0$) guarantee the existence of an
electric potential $\varphi$ that must satisfy (cf.,
\cite[Eqn.\,(4.12)]{virga:94})
\begin{equation}\label{eqn:Delta-varphi-plus}
  \Delta \varphi + \frac{\epsa}{\,\eperp}
  \div [ ( \nabla\varphi \cdot \nhat ) \nhat ] = 0 , \quad
  \bmE = - \nabla \varphi .
\end{equation}
The difference here is that the factor $\epsa/\eperp$ is $O(1)$ for
typical liquid-crystal materials, and the term involving $\nhat$ above
cannot be ignored---the liquid-crystal medium has a non-negligible
influence on the local electric field.
% We note that if the equation for the electric potential $\varphi$
% above were written in terms of the electric susceptibilities (related
% to the permittivites by $\epara=1+\chipara^{\text{e}}$,
% $\eperp=1+\chiperp^{\text{e}}$), then it would have the exact same
% form as the equation for the magnetic scalar potential $\phim$.
As an example, the values of these different factors for the material
5CB near $26^{\circ}$C are given by
\begin{equation*}
  \frac{\Delta\chi}{1+\chiperp} = 1.43 \times 10^{-6} , \quad
  \frac{\epsa}{\,\eperp} = 1.64 ,
\end{equation*}
using data given in \cite[Table\,D.3]{stewart:04}.

While there are other ways to see this, the above makes it clear that
for the systems that are of interest to us, the magnetic field can be
treated as uniform, unaffected by the medium (a true external field),
while the electric field should be viewed as coupled to the director
field.  These modeling assumptions lead to
\begin{equation*}
  \bmB \cdot \bmH = \chia ( \bmH\cdot\nhat )^2 + \text{const} , \quad
  \chia := \mu_0 \Delta \chi , \qquad
  \bmD \cdot \bmE = \epstensor(\nhat) \nabla\varphi \cdot \nabla\varphi ,
\end{equation*}
and the free energies for the magnetic field and electric field cases
have the forms
\begin{equation}\label{eqn:FH-FE}
  \begin{gathered}
    \calFH[\nhat] = \int_\Omega \Bigl[ \We(\nhat,\nabla\nhat) -
    \frac12 \chia (\bmH\cdot\nhat)^2 \Bigr] \, \td V , \\
    \calFE[\nhat,\varphi] = \int_\Omega \Bigl[ \We(\nhat,\nabla\nhat) -
    \frac12 \epstensor(\nhat) \nabla\varphi\cdot\nabla\varphi \Bigr]
    \, \td V .
  \end{gathered}
\end{equation}
% with the dielectric tensor $\epstensor$ as in \eqref{eqn:eps-tensor}.
% Most liquid-crystal materials (though not all) have $\chia>0$; most
% also have $\epsa>0$, though $\epsa<0$ is not uncommon.
Both $\chia$ and $\epsa$ can be \emph{positive} or \emph{negative}.

Globally stable solutions for a system with a magnetic field satisfy
\begin{equation*}
  \calFH[\nhat^*] = \min_{\nhat} \calF[\nhat] ,
\end{equation*}
subject to regularity assumptions, appropriate boundary conditions,
and the pointwise constraint $|\nhat|=1$.  The dielectric tensor
$\epstensor$ is real, symmetric, and \emph{positive definite}; so for
a system with an electric field, globally stable solution pairs are
characterized by
\begin{equation*}
  \calFE[\nhat^*\!\!,\varphi^*] =
  \min_{\nhat} \max_{\varphi} \calFE[\nhat,\varphi] ,
\end{equation*}
subject to appropriate auxiliary conditions on both $\nhat$ and
$\varphi$.  We note that this ``minimax'' problem can be written in an
equivalent way as a constrained minimization problem:
\begin{equation*}
  \calFE[\nhat^*\!\!,\varphi^*] =
  \min_{\nhat} \calFE[\nhat,\varphi] , ~ \text{subject to} \,
  \div [ \epstensor(\nhat) \nabla \varphi ] = 0 .
\end{equation*}
The constraint, which results from the maximization of $\calFE$ with
respect to $\varphi$ with a fixed director field $\nhat$, is
equivalent to the electrostatic equation $\div\bmD=0$ (another way of
writing \eqref{eqn:Delta-varphi-plus}).  In this setting, the electric
potential field can be thought of as ``slaved'' to the director field,
uniquely determined for any given $\nhat$ on $\Omega$.  The problem
for the globally stable solution in the presence of an electric field
can then be formulated in various ways: as a minimax problem for
$\nhat$ and $\varphi$, as a constrained minimization problem for
$\nhat$ (with $\varphi$ satisfying a partial differential equation
constraint), or in terms of a ``least free energy principle'' (in the
sense that a globally stable solution pair $(\nhat^*\!\!,\varphi^*)$
has the \emph{lowest} free energy among all \emph{equilibrium pairs}).

Most of the formulations in this section can be found in standard
references.  The Frank elastic model is well established, well
studied, and well presented in \cite{chandrasekhar:92,
  degennes:prost:93,virga:94,stewart:04}, as are the magnetic-field
\Freed\ transitions.  Some references take the approach that treating
the electric field in a liquid-crystal medium as uniform is a
reasonable approximation, based upon the assumption that
$\epsa/\eperp\ll1$.  This is how it is viewed in
\cite[Sec.\,4.1]{virga:94}, sometimes referred to in the literature as
``the magnetic approximation.''  As we have seen, this assumption is
not always valid.  Other references acknowledge that a nonuniform
director field can cause a nonuniform electric field, but they
maintain that \Freed\ thresholds are not affected by this.  This is
the case in \cite[Sec.\,IV.1]{pikin:91},
\cite[Sec.\,3.3.1]{degennes:prost:93}, and
\cite[Sec.\,3.5]{stewart:04}.  Only in
\cite[Sec.\,3.4.1]{chandrasekhar:92} does one find mention of the fact
that the analogy between electric and magnetic fields breaks down
``when the dielectric anisotropy is very large,'' allowing for a
first-order electric-field \Freed\ transition ``in certain
geometries'' (referencing \cite{arakelyan:karayan:chilingaryan:84,
  frisken:palffy:89,kini:90}).  Nothing is said there about the
formulas for the local instability thresholds.  We note that having a
large dielectric anisotropy is a desirable feature of a material for
most applications of liquid crystals.

\section{Magnetic-field \Freed\ transitions}

\label{sec:H-field-transitions}

\begin{figure}
  \centering
  \includegraphics[width=.32\textwidth]{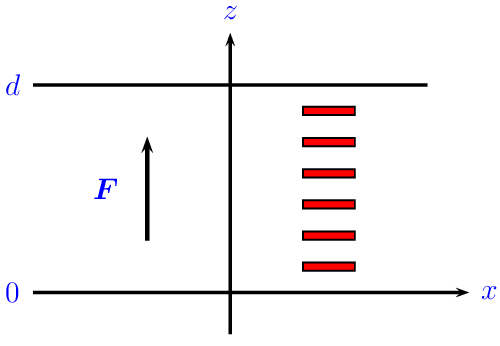}
  \includegraphics[width=.32\textwidth]{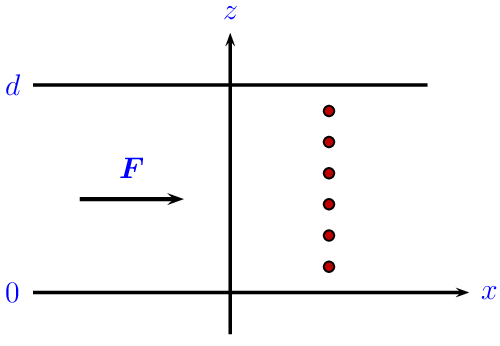}
  \includegraphics[width=.32\textwidth]{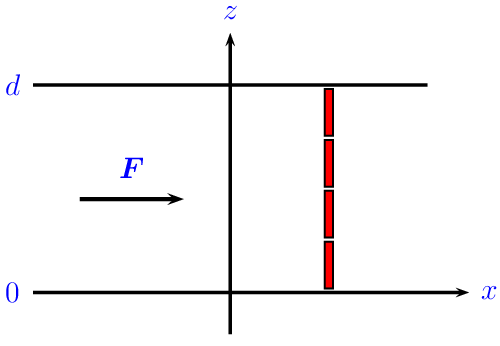}
  \caption{Geometries of the three classical \Freed\ transitions in
    \emph{magnetic} or \emph{electric} fields ($\bmF=\bmH$ or
    $\bmF=\bmE$) for a material with a \emph{positive} anisotropy
    ($\chia>0$ or $\epsa>0$): ``splay'' (left, director constrained to
    $x$-$z$ tilt plane), ``twist'' (center, director constrained to
    $x$-$y$ twist plane), ``bend'' (right, director constrained to
    $x$-$z$ tilt plane).} \label{fig:chia-pos-geoms}
\end{figure}

\begin{figure}
  \centering
  \hfill
  \includegraphics[width=.32\textwidth]{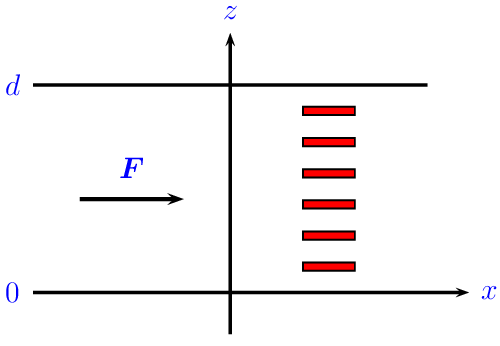}
  \hfill
  \includegraphics[width=.32\textwidth]{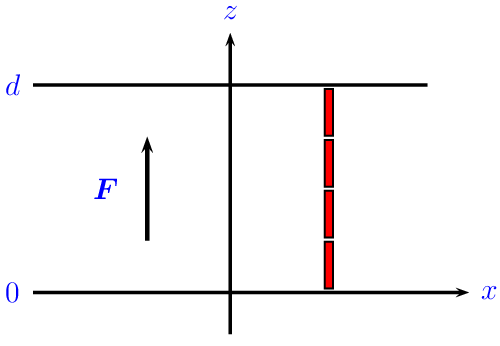}
  \hspace*{\fill}
  \caption{Geometries of classical \Freed\ transitions in
    \emph{magnetic} or \emph{electric} fields ($\bmF=\bmH$ or
    $\bmF=\bmE$) for materials of \emph{negative} anisotropy
    ($\chia<0$ or $\epsa<0$): ``splay'' (left, director constrained to
    $x$-$z$ tilt plane), ``twist'' (left, director constrained to
    $x$-$y$ twist plane), ``bend'' (right, director constrained to
    $x$-$z$ tilt plane).}
  \label{fig:chia-neg-geoms}
\end{figure}

We summarize results for the classical magnetic-field \Freed\
transitions from a variational point of view, as found, for example,
in \cite{virga:94,stewart:04}.  There are six cases: three
corresponding to materials with positive magnetic anisotropy
($\chia>0$), three for materials with $\chia<0$.  The geometries of
the classical transitions (with either magnetic or electric fields)
are depicted in Figures\,\ref{fig:chia-pos-geoms} and
\ref{fig:chia-neg-geoms} for reference.  These illustrate the
orientations of the ground-state director fields and the external
magnetic or electric fields, as well as the coordinate system to be
used here.  The anchoring conditions are assumed to be infinitely
strong and the lateral dimensions sufficiently large compared to the
cell gap $d$ such that the director field can be assumed to be uniform
in those directions (a function of $z$ only).  In each case, the
distortion of the director field is assumed to be restricted to a
plane: the $x$-$z$ tilt plane for the splay and bend transitions, the
$x$-$y$ twist plane for the twist transitions.  The restriction of the
director distortion to a plane is a bit of an idealization, as some
additional mechanism (such as an additional external field) could be
required to enforce this in some cases.

By virtue of the fact that the director field is assumed to be planar
in each geometry, $\nhat$ can be represented in terms of a single tilt
or twist angle, and such a representation is convenient in analyzing
the classical \Freed\ transitions.  For more general liquid-crystal
systems (and for the more general theory of field-induced
instabilities developed in \cite{gartland:21}), the vector
representation of $\nhat$ is advantageous.  An equilibrium director
field for a system such as ours governed by a free energy $\calFH$ of
the form in \eqref{eqn:FH-FE} can be characterized as a solution of a
system of Euler-Lagrange equations
% \begin{equation}\label{eqn:H-Euler-Lagrange}
%   - \frac{\upartial}{\upartial x_j}
%   \Bigl( \frac{\upartial\We}{\upartial n_{i,j}} \Bigr) +
%   \frac{\upartial\We}{\upartial n_i} = \lambda n_i +
%   \chia (H_j n_j) H_i , \text{ in } \Omega ,
% \end{equation}
\begin{equation*}%\label{eqn:H-Euler-Lagrange}
  - \div \Bigl( \frac{\upartial\We}{\upartial\nabla\nhat} \Bigr) +
  \frac{\upartial\We}{\upartial\nhat} = \lambda \nhat +
  \chia (\bmH\cdot\nhat) \bmH , \text{ in } \Omega ,
\end{equation*}
subject to appropriate boundary conditions and the unit-length
constraint $|\nhat|=1$ (which must hold at each point in the domain
$\Omega$).  A certain amount of regularity of $\nhat$ is also assumed.
The scalar field $\lambda$ is a Lagrange-multiplier field associated
with the unit-length constraint and is given by
% \begin{equation*}
%   \lambda = \Bigl[ - \frac{\upartial}{\upartial x_j}
%   \Bigl( \frac{\upartial\We}{\upartial n_{i,j}} \Bigr) n_i +
%   \frac{\upartial\We}{\upartial n_i} n_i \Bigr] - \chia (H_j n_j)^2 ,
% \end{equation*}
\begin{equation*}
  \lambda = \Bigl[ - \div \Bigl(
  \frac{\upartial\We}{\upartial\nabla\nhat} \Bigr) +
  \frac{\upartial\We}{\upartial\nhat} \Bigr] \cdot \nhat -
  \chia (\bmH\cdot\nhat)^2 ,
\end{equation*}
with the expression above evaluated at the equilibrium director field.
Note that $\lambda$ is a scalar \emph{field}; though in simple cases,
it could be a constant.  In particular, $\lambda$ is constant for the
ground states of all six of the basic \Freed\ geometries, and
$\lambda=0$ for the three transitions with $\chia>0$---this follows
from the vanishing of the bracketed term above for any spatially
uniform vector field $\nhat$ (with $\We$ as in \eqref{eqn:We}) and the
fact that $\bmH\cdot\nhat = 0$ for the geometries with $\chia>0$
(those in Figure\,\ref{fig:chia-pos-geoms}).  These conditions hold as
well for the analogous electric-field cases.
% The Lagrange multiplier can be viewed as the response by the system (a
% kind of generalized force) that is needed to preserve the constraint
% $|\nhat|=1$, which other influences (distortional elasticity and/or
% external field) are trying to cause to be violated.  It is the energy
% per unit volume associated with moving off the constraint manifold.
% The Lagrange multiplier can be viewed as the energy cost per unit
% volume associated with violating the constraint $|\nhat|=1$ at a
% point, the response by the system that is needed to preserve the
% constraint, which other influences (distortional elasticity and/or
% external fields) are trying to cause to be violated.

For an equilibrium director field $\nhat_0$ to be \emph{locally
  stable} (metastable) in general, its free energy should lie at the
bottom of a well on the free-energy surface.  This can be determined
by evaluating the free energy on admissible director fields close to
$\nhat_0$.  Such fields can be produced, for example, in the form
\begin{equation*}
  \neps = \frac{\nhat_0+\epsilon\bmv}{|\nhat_0+\epsilon\bmv|} ,
\end{equation*}
as is used in \cite[Sec.\,3.5]{virga:94} and
\cite{kinderlehrer:ou:92}.  The fields $\neps$ are normalized so that
$|\neps|=1$ on $\Omega$ (for any $\bmv$, with $|\epsilon|$
sufficiently small).  The perturbative field $\bmv$ must also be such
that $\neps$ satisfies the boundary conditions that $\nhat$ must
satisfy.  In a general setting, these conditions could include strong
anchoring, weak anchoring, periodic boundary conditions, etc.,\ and an
account along those lines is given in \cite{gartland:21}.  Here, for
simplicity, we assume that $\nhat$ must satisfy \emph{strong
  anchoring} conditions on the entire boundary of $\Omega$ (denoted
$\partial\Omega$).  Thus for $\neps$ to be admissible, $\bmv$ must
\emph{vanish} on $\partial\Omega$.  Expanding for small $\epsilon$
gives
\begin{equation}\label{eqn:neps-u-P-v}
  \neps = \nhat_0 + \epsilon \bmu + O(\epsilon^2) , \quad
  \bmu = \bfP(\nhat_0) \bmv , \quad
  \bfP(\nhat) := \bfI - \nhat \otimes \nhat .
\end{equation}
The linear operator $\bfP(\nhat)$ gives the projection transverse to
$\nhat$: $\bfP(\nhat)\bmv \perp \nhat$ (cf.,
\cite[Sec.\,3.2]{virga:94}).  If $\lambda_0$ is the Lagrange
multiplier field associated with $\nhat_0$, then the expansion of the
free energy of $\neps$ can be expressed
\begin{equation}\label{eqn:FH-neps}
  \calFH[\neps] = \calFH[\nhat_0] + \frac12 \epsilon^2 \Bigl[
  \udelta^2 \! \calFH[\nhat_0](\bmu) -
  \int_\Omega \lambda_0 | \bmu |^2 \, \td V \Bigr] + o(\epsilon^2) ,
\end{equation}
with $\bmu=\bfP(\nhat_0)\bmv$ as above.  Here $\udelta^2\!\calFH$
denotes the \emph{second variation} of the functional $\calFH$ and has
the form
\begin{multline*}
  \udelta^2 \! \calFH[\nhat](\bmv) = \int_\Omega \Bigl(
  \frac{\upartial^2 W}{\upartial n_i \upartial n_k} v_i v_k +
  2 \frac{\upartial^2 W}{\upartial n_i \upartial n_{k,l}} v_i v_{k,l} +
  \frac{\upartial^2 W}{\upartial n_{i,j} \upartial n_{k,l}} v_{i,j} v_{k,l}
  \Bigr) \td V , \\
  W = \We - \frac12 \chia (\bmH\cdot\nhat)^2 .
\end{multline*}
A derivation of \eqref{eqn:FH-neps} can be found in
\cite{gartland:21}.  A term of first order in $\epsilon$ is not
present in \eqref{eqn:FH-neps} because of the equilibrium conditions
satisfied by $\nhat_0$ and $\lambda_0$.  The integral with the
Lagrange-multiplier field $\lambda_0$ compensates for the curvature of
the constraint manifold $|\nhat_0|=1$, which the perturbation $\bmu$
enforces only to first order.  The necessary condition for local
stability of the equilibrium director field $\nhat_0$, with associated
Lagrange-multiplier field $\lambda_0$, can then be expressed
\begin{equation}\label{eqn:H-local-stability}
  \udelta^2 \! \calFH[\nhat_0](\bmu) -
  \int_\Omega \lambda_0 |\bmu|^2 \, \td V \ge 0 ,
\end{equation}
and this must hold for all smooth test fields $\bmu$ that are
\emph{transverse} to $\nhat_0$ ($\nhat_0\cdot\bmu=0$ on $\Omega$) and
\emph{vanish} on the boundary $\partial\Omega$.  This form of local
stability assessment follows the approaches of
\cite{kinderlehrer:ou:92} and \cite{rosso:virga:kralj:04}.

As an illustration, consider the case of the splay-\Freed\ transition
with $\chia<0$ (Figure\,\ref{fig:chia-neg-geoms} left).  For this
geometry,
\begin{equation*}
  \nhat = n_x(z) \ex + n_z(z) \ez ,
\end{equation*}
the free energy (per unit cross-sectional area) has the form
\begin{equation*}%\label{eqn:FH-splay-bend}
  \calFH[\nhat] = \frac12 \int_0^d
  \bigl( K_1 n_{z,z}^2 + K_3 n_{x,z}^2 - \chia H^2 n_x^2 \bigr)
  \, \td z ,
\end{equation*}
and the equilibrium Euler-Lagrange system is given by
% \begin{equation}\label{eqn:EL-splay-bend}
%   \begin{gathered}
%     K_3 n_{x,zz} + \bigl( \chia H^2 + \lambda \bigr) n_x = 0 , \quad
%     K_1 n_{z,zz} + \lambda n_z = 0 , \quad 0 < z < d , \\
%     n_x(0) = n_x(d) = 1 , \quad n_z(0) = n_z(d) = 0 , \quad
%     n_x^2 + n_z^2 = 1 .
%   \end{gathered}
% \end{equation}
\begin{equation*}%\label{eqn:EL-splay-bend}
  \begin{gathered}
    K_3 n_{x,zz} + \bigl( \chia H^2 + \lambda \bigr) n_x = 0 , \quad
    K_1 n_{z,zz} + \lambda n_z = 0 , \quad
    n_x^2 + n_z^2 = 1 , \quad 0 < z < d , \\
    n_x(0) = n_x(d) = 1 , \quad n_z(0) = n_z(d) = 0 .
  \end{gathered}
\end{equation*}
The ground-state equilibrium solution of the above is given by
\begin{equation*}
  \nhat_0 = \ex ~~ ( n_x = 1, n_z = 0 ) , \quad
  \lambda_0 = - \chia H^2 .
\end{equation*}
Boundary conditions with $n_x=-1$ and a ground state of $\nhat_0=-\ex$
would work equally well.  Test fields $\bmu$ in
\eqref{eqn:H-local-stability} must here be $x$-$z$ planar and
orthogonal to $\nhat_0$ on $0<z<d$, and they must vanish at $z=0$ and
$z=d$.  Thus they must be of the form
\begin{equation*}
  \bmu = w(z) \ez , \quad w(0) = w(d) = 0 .
\end{equation*}
The stability condition \eqref{eqn:H-local-stability} then becomes
\begin{equation*}%\label{eqn:H-splay-neg-stability}
  \udelta^2\!\calFH[\nhat_0](\bmu) -
  \int_0^d \lambda_0 |\bmu|^2 \, \td z =
  K_1 \int_0^d \wz^2 \, \td z +
  \chia H^2 \int_0^d w^2 \, \td z \ge 0 ,
\end{equation*}
which implies
\begin{equation*}
  - \frac{\chia H^2}{K_1} \le
  \frac{\int_0^d \wz^2 \, \td z}{\int_0^d w^2 \, \td z} ,
\end{equation*}
and this must hold for all smooth functions $w$ that vanish at $z=0$
and $z=d$.  The minimum value of the expression on the right-hand side
above over all such functions is given by
\begin{equation}\label{eqn:Rayleigh-quotient}
  \min_{w(0)=w(d)=0}
  \frac{\int_0^d \wz^2 \, \td z}{\int_0^d w^2 \, \td z} =
  \frac{\,\pi^2}{\,d^2} , ~~ \text{ attained by }
  w = \sin \frac{\pi z}{d} .
\end{equation}
This can most readily be seen by using a modal expansion
$w = \sum b_n \sin( n\pi z/d)$.  The minimal value and attaining
function above represent the minimum eigenvalue and associated
eigenfunction of the eigenvalue problem
\begin{equation*}
  \frac{\td^2w}{\td z^2} + \lambda w = 0 , \quad w(0) = w(d) = 0 .
\end{equation*}
It follows that in order for the ground state director field
$\nhat_0=\pm\ex$ to be locally stable, the strength of the magnetic
field must satisfy
\begin{equation*}
  H \le \frac{\pi}{d} \sqrt{\frac{K_1}{-\chia}} ,
\end{equation*}
which is the correct instability threshold.

As modeled above, the local stability assessments for all six
magnetic-field \Freed\ transitions reduce to similar inequalities and
ultimately rest upon the minimum-eigenvalue formula
\eqref{eqn:Rayleigh-quotient}, giving the classical threshold formulas
for the splay, twist, and bend transitions (valid for both $\chia>0$
and $\chia<0$):
\begin{equation*}
  \Hc = \frac{\pi}{d} \sqrt{\frac{K_i}{|\chia|}} , \quad i = 1, 2, 3 .
\end{equation*}
See \cite[Sec.\,IV.1]{pikin:91}, \cite[Sec.\,3.4.1]{chandrasekhar:92},
\cite[Sec.\,3.2.3]{degennes:prost:93}, \cite[Sec.\,4.2]{virga:94},
\cite[Sec.\,3.4.1]{stewart:04}.
% For such simple geometries and systems, the stability inequalities
% that result from \eqref{eqn:H-local-stability} are quite simple as
% well.
Rather than appeal to a general characterization of stability as in
\eqref{eqn:H-local-stability}, most references analyze individual
\Freed\ transitions by direct consideration of appropriate free
energies (expressed in terms of angle representations of $\nhat$),
using first integrals obtained from the associated Euler-Lagrange
equations.  The textbook analyses also go a bit further and show that
the classical magnetic-field \Freed\ transitions are always
\emph{second order}, information that is not provided by
\eqref{eqn:H-local-stability}.  The advantage of \eqref{eqn:FH-neps}
and \eqref{eqn:H-local-stability} for our purposes is the way that
they generalize to electric-field-induced instabilities, which are to
follow.  The example discussed above also illustrates the role that
the Lagrange-multiplier field associated with an equilibrium director
field can play in the assessment of the local stability of that
director field.

% Most references analyze magnetic-field \Freed\ transitions only for
% the cases with $\chia>0$.  This can be attributed, at least in part,
% to the fact that results for systems with $\chia<0$ can be obtained
% from those for systems with $\chia>0$ by simple substitutions of
% parameters.

% For example, the magnetic-field bend-\Freed\ transition with $\chia>0$
% (Figure\,\ref{fig:chia-pos-geoms} right) has the same free energy and
% Euler-Lagrange equations as those for the splay transition with
% $\chia<0$ analyzed above (viz., \eqref{eqn:FH-splay-bend} and
% \eqref{eqn:EL-splay-bend}).  The two systems differ only in their
% ground states and admissible variations:
% \begin{gather*}
%   \nhat_0 = \pm \ex , \quad \lambda_0 = - \chia H^2 , \quad
%   \bmu = w(z) \ez \quad (\text{splay transition}, \chia<0) , \\
%   \nhat_0 = \pm \ez , \quad \lambda_0 = 0 , \quad
%   \bmu = u(z) \ex \quad (\text{bend transition}, \chia>0) .
% \end{gather*}
% The condition that results for local stability of the ground state in
% the bend geometry is
% \begin{equation*}
%   K_3 \int_0^d \wz^2 \, \td z - \chia H^2 \int_0^d w^2 \, \td z \ge 0 ,
% \end{equation*}
% which is the same as \eqref{eqn:H-splay-neg-stability} with the
% substitutions $K_1 \leftrightarrow K_3$ and $\chia \leftrightarrow
% -\chia$.  The critical field
% \begin{equation*}
%   \Hc = \frac{\pi}{d} \sqrt{\frac{K_3}{\chia}}
% \end{equation*}
% is thus obtained.

\section{Models of electric-field \Freed\ transitions}

\label{sec:models-of-E-field-transitions}

Before discussing the characterization of local stability for
equilibria with electric fields (generalizing \eqref{eqn:FH-neps} and
\eqref{eqn:H-local-stability}), we first illustrate the anomalous
behavior of one of these transitions by direct analysis of its
Euler-Lagrange equations.  For now we consider only materials with
$\epsa>0$---results for materials with $\epsa<0$ can be obtained
directly from these by simple substitutions of parameters.  Thus we
consider the three classical \Freed\ geometries in
Figure\,\ref{fig:chia-pos-geoms}, but with electric fields instead of
magnetic fields.  The electric fields are created by electrodes, which
in the case of the splay geometry (Figure\,\ref{fig:chia-pos-geoms}
left) are on the top and bottom of the cell.  For the twist and bend
geometries (Figure\,\ref{fig:chia-pos-geoms} center, right), the
electric field needs to be in the plane of the film, and so the
electrodes must be placed on the left and right ends (across the wide
dimension of the cell).  We refer to the first case as ``cross plane''
and the latter cases as ``in plane.''  The two cases require somewhat
different handling.

\subsection{Cross-plane electric field}

The electric-field splay-\Freed\ transition was analyzed in
\cite{gruler:scheffer:meier:72} and \cite{deuling:72} taking into
account the coupling between the director field and the electric
field.  The electric field is confined to the cell, with a potential
difference $V$ between the top and bottom electrodes.  By assumption,
$\nhat = n_x(z)\ex+n_z(z)\ez$ and $\varphi=\varphi(z)$.  The relations
$\bmE = -\nabla\varphi$ and $\div\bmD=0$ then give
\begin{equation*}
  \bmE = - \varphiz \ez , \quad D_z = \text{const} .
\end{equation*}
Using these together with $\varphi(d)-\varphi(0)=V$, one can deduce
the constant value of $D_z$ and a non-local expression for $\bmE$ in
terms of $\nhat$ (cf., \cite[Eqns.\,(3.220) and (3.221)]{stewart:04}):
\begin{equation}\label{eqn:Dz-E-splay}
  D_z = - \eps_0 V \biggl[ \int_0^d \!
  \frac{\td z}{\eperp n_x^2 + \epara n_z^2} \biggr]^{-1} , \quad
  \bmE = - V \biggl[ \int_0^d \!
  \frac{\td z}{\eperp n_x^2 + \epara n_z^2} \biggr]^{-1} \!\!\!
  \frac1{\eperp n_x^2 + \epara n_z^2} \, \ez .
\end{equation}
These enable one to write the free energy $\calFE$ for this system in
the form of a \emph{reduced} functional of $\nhat$ only:
\begin{equation}\label{eqn:Fn-splay}
  \calF[\nhat] = \frac12 \int_0^d \!
  \bigl( K_1 n_{z,z}^2 + K_3 n_{x,z}^2 \bigr) \, \td z -
  \frac12 \eps_0 V^2 \biggl[ \int_0^d
  \frac{\td z}{\eperp n_x^2 + \epara n_z^2} \biggr]^{-1} .
\end{equation}

The analyses of \cite{gruler:scheffer:meier:72} and \cite{deuling:72}
were based upon studies of this functional and concluded that the
critical voltage at which the uniform horizontal ground state of the
director field becomes unstable is given by
\begin{equation*}
  \Vc = \pi \sqrt{\frac{K_1}{\eps_0\epsa}}
\end{equation*}
and that the transition is \emph{second order} for any $K_1, K_3 > 0$
and $\epara>\eperp>0$.  This instability follows the magnetic-field
analogy:
\begin{equation*}
  \Ec = \frac{\,\Vc}{d} = \frac{\pi}{d} \sqrt{\frac{K_1}{\eps_0\epsa}}
  ~ \leftrightarrow ~
  \Hc = \frac{\pi}{d} \sqrt{\frac{K_1}{\chia}} .
\end{equation*}
The same results can be obtained by analyzing the free energy modeled
as a functional of \emph{both} $\nhat$ and $\varphi$, which follows
from the expression for $\calFE$ in \eqref{eqn:FH-FE}:
\begin{equation*}%\label{eqn:FnU}
  \calF[\nhat,\varphi] = \frac12 \int_0^d
  \bigl[ K_1 n_{z,z}^2 + K_3 n_{x,z}^2 - \eps_0
  \bigl( \eperp n_x^2 + \epara n_z^2 \bigr)
  \varphiz^2 \bigr] \, \td z .
\end{equation*}
While this will now lead to a \emph{coupled} set of equilibrium
equations for $\nhat$ and $\varphi$, the model has the advantage of
being \emph{local}, whereas the Euler-Lagrange equations associated
with \eqref{eqn:Fn-splay} above are of integro-differential type (cf.,
\cite[Eqn.\,(3.226)]{stewart:04}).
% See Appendix~\ref{app:pert-anal-of-E-field-splay}.

\subsection{In-plane electric field}

\label{sec:in-plane-electric-field}

The electric-field twist and bend geometries both involve an electric
field oriented in the plane of the film.  This is an awkward geometry
for experiments (electrodes on the wide ends of the cell), as well as
for modeling (the electric field extending above and below the cell).
Assuming as before that the cell gap is small compared to the lateral
dimensions, boundary-layer theory suggests that the director field and
electric field can be well approximated by ``outer solutions'' of the
form $\nhat=\nhat(z)$ and $\bmE=\bmE(z)$ throughout most of the region
plus boundary-layer corrections, the influence of which is significant
only in narrow regions adjacent to the lateral boundaries.  We shall
thus model the systems with in-plane electric fields in terms of the
\emph{outer solutions} in this asymptotic regime, as was also done in
\cite{arakelyan:karayan:chilingaryan:84,frisken:palffy:89,frisken:palffy:89c}.
This modeling assumption was validated to a degree by numerical
experiments in \cite[\S4.3]{richards:06}.

We assume a simple experimental setup, with tall electrodes on the
left and right ends and with the regions above and below the cell
consisting of stratified layers of homogeneous materials (alignment
layers, polarizers, glass substrates, air, etc.), all with horizontal
interfaces.  This is the type of system that was used in the
experiments reported in \cite{frisken:palffy:89,frisken:palffy:89b}.
Other experimental setups are possible (interdigitated electrodes,
electrode strips as spacers), but it does not appear that they produce
as uniform an electric field in the sample.  Utilizing the assumptions
$\nhat=\nhat(z)$ and $\bmE=\bmE(z)$ in the basic electrostatic
equations $\curl\bmE=\bmzero$ and $\div\bmD=0$, together with
\emph{interface conditions} ($E_x$, $E_y$, $D_z$ continuous) and
\emph{far-field conditions} as $z\rightarrow\pm\infty$ ($E_x=E_0$,
$E_y=0$, $D_z=0$), one obtains an expression for the electric field
that is valid throughout the entire region between the electrodes
(inside the liquid-crystal cell, as well as in the extended regions
above and below it):
\begin{equation*}%\label{eqn:twist-bend-E}
  \bmE = E_0 \ex + E_z(z) \ez , \quad E_0 := - V / l ,
\end{equation*}
with $E_z=0$ outside $0\le z\le d$.  Here $E_0\ex$ is the uniform
electric field that one would have if there were no inhomogeneity in
the medium between the electrodes, with $V$ the potential difference
between the electrodes and $l$ the distance separating them.

Using the above expression for $\bmE$ in the relation $D_z=0$ (with
$\bmD$ as in \eqref{eqn:eps-tensor}) gives
\begin{equation*}
  \bigl[ \eperp \bigl( n_x^2 + n_y^2 \bigr) + \epara n_z^2 \bigr] E_z +
  \epsa n_x n_z E_0 = 0 ,
\end{equation*}
which can be solved for $E_z$ in $0<z<d$ in both the twist and bend
cases: in the \emph{twist} case,
\begin{subequations}
\begin{equation}\label{eqn:n-Ez-twist}
  \nhat = n_x\ex + n_y\ey ~ \Rightarrow ~ E_z = 0 ,
\end{equation}
in the \emph{bend} case,
\begin{equation}\label{eqn:n-Ez-bend}
  \nhat = n_x\ex + n_z\ez ~ \Rightarrow ~
  E_z = - \epsa E_0 \frac{n_x n_z}{\eperp n_x^2 + \epara n_z^2} .
\end{equation}
\end{subequations}
% From these follow expressions for $W_{\text{E}}$ for the two cases:
% \begin{equation*}%\label{eqn:twist-bend-WE}
%   W_{\text{E}} = - \frac12 \eps_0
%   \bigl( \epara n_x^2 + \eperp n_y^2 \bigr) E_0^2
%   ~~ (\text{twist}) , \quad
%   W_{\text{E}} = - \frac12 \eps_0
%   \frac{\eperp\epara}{\eperp n_x^2 + \epara n_z^2} E_0^2
%   ~~ (\text{bend}) .
% \end{equation*}
It can be shown that the expressions for the electric field given
above are precisely what one obtains if one formulates the full
coupled system for the director field $\nhat$ and electric potential
$\varphi$ (as functions of $x$ and $z$), rescales introducing the
small parameter $\eta:=d/l$ (where $d$ is the cell gap and $l$ the
distance between the electrodes), forms the problems for the outer
solutions $\nhat$ and $\varphi$ (by setting $\eta=0$), and solves the
resulting problem for $\varphi$ in terms of $\nhat$.  The expression
for $E_z$ in the bend case in \eqref{eqn:n-Ez-bend} is as found in
\cite{arakelyan:karayan:chilingaryan:84}.

% \begin{equation*}%\label{eqn:Fn-twist}
%   \calF[\nhat] = \frac12 \int_0^d \bigl[ K_2 (n_xn_{y,z}-n_yn_{x,z})^2 -
%   \eps_0 \bigl( \epara n_x^2 + \eperp n_y^2 \bigr) E_0^2 \bigr] \,
%   \td z .
% \end{equation*}

% \begin{equation*}%\label{eqn:Fn-bend}
%   \calF[\nhat] = \frac12 \int_0^d \Bigl[ K_1 n_{z,z}^2 + K_3 n_{x,z}^2 -
%   \eps_0 \frac{\eperp\epara}{\eperp n_x^2 + \epara n_z^2} E_0^2 \Bigr]
%   \td z .
% \end{equation*}

Using the expressions for $\nhat$ and $\bmE$ in \eqref{eqn:n-Ez-twist}
and \eqref{eqn:n-Ez-bend} above, one can write the electric-field free
energy $\calFE$ in \eqref{eqn:FH-FE} for the twist and bend cases in
the following forms:
\begin{subequations}
  \begin{gather}
    \calF[\nhat] = \frac12 \int_0^d \bigl[ K_2 (n_xn_{y,z}-n_yn_{x,z})^2 -
    \eps_0 \bigl( \epara n_x^2 + \eperp n_y^2 \bigr) E_0^2 \bigr] \,
    \td z ~~ (\text{twist geometry}) , \label{eqn:Fn-twist} \\
    \calF[\nhat] = \frac12 \int_0^d \Bigl[ K_1 n_{z,z}^2 + K_3 n_{x,z}^2 -
    \eps_0 \frac{\eperp\epara}{\eperp n_x^2 + \epara n_z^2} E_0^2 \Bigr]
    \td z ~~ (\text{bend geometry}) . \label{eqn:Fn-bend}
  \end{gather}
\end{subequations}
Thus, modeling the cases with in-plane electric fields in terms of the
``outer solutions'' ($\nhat=\nhat(z)$, $\bmE=\bmE(z)$) gives rise to
reduced models $\calF=\calF[\nhat]$ that are \emph{local}, in contrast
to the non-local reduced model that one obtains for the electric-field
splay-\Freed\ transition.  The reality of the matter is that the
electric field at a point in $\Omega$ depends on the director field
everywhere; however, in this geometry and asymptotic regime, the
non-local contribution only serves as a small correction associated
with the boundary-layer functions.  The modeling of the electric field
here is essentially equivalent to that done in
\cite{arakelyan:karayan:chilingaryan:84}.  If one wishes to model the
electric-field twist and bend transitions in a way that is as close as
possible to the way that the corresponding magnetic-field transitions
are modeled, then one is led to make the assumptions made here
(i.e., $\nhat=\nhat(z)$, $\bmE=\bmE(z)$).

\subsection{Discussion}

The basic assumption that the width of the cell is much smaller than
the lateral dimensions is taken as a justification for assuming that
equilibrium fields satisfy $\nhat=\nhat(z)$ and $\bmE=\bmE(z)$ (to a
good degree of approximation).  This underlies the modeling for all
three electric-field \Freed\ transitions, though it leads to different
consequences in each case.  In the splay geometry (the only case with
a cross-plane electric field), the assumed functional dependence of
$\nhat$ and $\bmE$ only on $z$ enables one to derive the explicit
formula for the electric field in terms of the director field given in
\eqref{eqn:Dz-E-splay}.  This expression is non-local, because it
depends on $\int_0^d (\eperp n_x^2 + \epara n_z^2)^{-1}\td z$, but it
leads to the reduced free-energy functional \eqref{eqn:Fn-splay},
which is amenable to analysis.

For the geometries with in-plane electric fields (twist and bend), the
same assumptions enable one to deduce explicit formulas for the
electric field in terms of the director field that are
local---$\bmE(z)$ depends only on the value of $\nhat$ at the point
$z$.  In the twist case, the electric field is in fact \emph{uniform}
throughout the strip between the electrodes (which are viewed as being
infinitely tall):
\begin{equation*}
  \bmE = E_0 \ex , \quad - \infty < z < \infty .
\end{equation*}
While in the bend case, the electric field is uniform \emph{outside}
the cell, but inhomogeneity of the director field induces
nonuniformity of the electric field \emph{inside} the cell:
\begin{equation}\label{eqn:E-Ez-bend}
  \bmE = E_0 \ex + E_z(z) \ez , \quad
  E_z = - \epsa E_0 \frac{n_x n_z}{\eperp n_x^2 + \epara n_z^2} , \quad
  0 < z < d .
\end{equation}
In all three cases, the formulas for the electric fields in terms of
the director fields enable one to write the free energies as reduced
functionals of $\nhat$ only: \eqref{eqn:Fn-splay},
\eqref{eqn:Fn-twist}, and \eqref{eqn:Fn-bend}.

It is not difficult to see why these three cases differ.  In the splay
geometry, the electrodes are on the top and bottom, and the electric
field is confined to the cell.  The value of $D_z$ is constant across
the cell but \emph{nonzero}, by virtue of the surface charge densities
on the electrodes---with a charge density of $\sigmaf$ on $z=d$ and
$-\sigmaf$ on $z=0$, we have $D_z=-\sigmaf$.  To express this constant
in terms of $\nhat$ requires an integration $\int_0^d \td z$ and
results in the non-local expression in \eqref{eqn:Dz-E-splay}.

The twist and bend geometries both involve in-plane electric fields,
and in both cases, the electric fields extend beyond the cell
containing the liquid crystal to the whole region between the vertical
electrodes, $-\infty<z<\infty$.  For both cases, $D_z$ is constant
throughout this whole region, and this constant is in fact
\emph{zero}, by virtue of the far-field conditions
($\bmE\rightarrow E_0\ex$, as $z\rightarrow\pm\infty$).  The relation
$D_z=0$ can be solved for $E_z$ in terms of $\nhat$, resulting in the
local expressions in \eqref{eqn:n-Ez-twist} and \eqref{eqn:n-Ez-bend}.
In both the twist geometry and the bend geometry, the equilibrium
electric field is always uniform \emph{outside} the cell, as well as
\emph{inside} the cell in the \emph{ground state}: $\bmE=E_0\ex$.  The
equilibrium electric field maintains this same uniform structure in
the cell for any \emph{twist} deformation, while \emph{bend}
distortions of the director cause the electric field in the cell to
develop inhomogeneity in the form \eqref{eqn:E-Ez-bend}.
% \begin{equation*}
%   \bmE = E_0 \ex + E_z(z) \ez ,
% \end{equation*}
% with $E_z$ as given in \eqref{eqn:E-Ez-bend}.
The reason why $E_z=0$ in the cell for twist distortions but
$E_z\not=0$ for bend distortions can be understood in terms of the
different nature of the \emph{dielectric polarization} in the two
cases.

There are two sources of contributions to the local electric field in
the systems under consideration here: free charges on the surface of
the electrodes and induced polarization in the dielectric medium.  The
electrodes are essentially accounted for in the horizontal component
$E_0\ex$, the uniform electric field that one would have if the medium
were homogeneous.  The electric field associated with a polarization
field $\bmP$ in a volume $\Omega$ is the same as the field that would
be created by an effective volume charge density $\rhoP$ in $\Omega$
and surface charge density $\sigmaP$ on the boundary $\partial\Omega$
given by
\begin{equation*}
  \rhoP = - \div \bmP , \quad \sigmaP = \bmP\cdot\nuhat ,
\end{equation*}
where $\nuhat$ is the outward unit normal on $\partial\Omega$.
%\cite[Secs.\,4-2,\,4-3]{reitz:milford:christy:09}.
Under our modeling assumptions (transversely isotropic material,
linear dielectric regime), the induced polarization in the
liquid-crystal layer is given by
\begin{equation}\label{eqn:P-chie}
  \bmP = \chieperp \bmE + \Delta\chie (\bmE\cdot\nhat) \nhat , \quad
  \Delta \chie := \chiepara - \chieperp ,
\end{equation}
where $\chiepara$ and $\chieperp$ are the electric susceptibilities
parallel to $\nhat$ and perpendicular to $\nhat$.  These are related
to the electric permittivities through $\bmD=\eps_0\bmE+\bmP$ and
\eqref{eqn:eps-tensor}:
\begin{equation}\label{eqn:eps-vs-chi}
  \eps_0 \eperp = \eps_0 + \chieperp , \quad
  \eps_0 \epara = \eps_0 + \chiepara , \quad
  \eps_0 \epsa = \Delta \chie .
\end{equation}
In the asymptotic regime of the outer solutions, $\bmP=\bmP(z)$, and
all interfaces are horizontal ($\nuhat=\pm\ez$).  It follows that
\begin{equation*}
  \rhoP = - P_{z,z} , \quad \sigmaP = \pm P_z .
\end{equation*}
The $z$ component of $\bmP$ is identically \emph{zero} throughout the
region between the electrodes \emph{exterior} to the cell.  Thus, in
order for a polarization field $\bmP$ in the liquid-crystal cell (in
our asymptotic regime) to produce an electric field, it must be such
that $P_z$ has a non-vanishing $z$ derivative in $0<z<d$ or
$P_z\not=0$ on $z=0+$ or $z=d-$.

If we contrast twist distortions with bend distortions in our geometry
($\nhat=n_x(z)\ex+n_y(z)\ey$ versus $\nhat=n_x(z)\ex+n_z(z)\ez$),
assuming for the moment a uniform electric field $\bmE=E_0\ex$, we
obtain from \eqref{eqn:P-chie}
\begin{equation*}
  P_z^{\text{twist}} = 0 , \quad
  P_z^{\text{bend}} = \Delta\chie E_0 n_x n_z , \quad 0 < z < d .
\end{equation*}
Thus for twist distortions, $\rhoP=0$ and $\sigmaP=0$.  Twist
distortions give rise to changes in the $x$ and $y$ components of
$\bmP$, but these are \emph{uniform} in the $x$-$y$ plane under our
assumptions and do not contribute to $\div\bmP$ ($P_{x,x}=P_{y,y}=0$).
Even though the twist-distorted director field has a nonzero $z$
derivative, $P_z$ is constant (zero) and does not.  Thus a uniform
electric field $\bmE=E_0\ex$ is in electrostatic equilibrium with any
twist configuration.

For bend distortions, $n_x=0$ on $z=0$ and $z=d$; so $\sigmaP=0$, as
with twist distortions.  However,
\begin{equation*}
  P_{z,z}^{\text{bend}} =
  \Delta\chie E_0 \frac{\upartial}{\upartial z} (n_xn_z) \not=0 ,
\end{equation*}
by virtue of the variation of $\nhat$ in the $z$ direction.  Thus
$\rhoP\not=0$, and there is an associated \emph{effective charge
  density}, which would give rise to additional contributions to
$\bmE$.  A uniform electric field $\bmE=E_0\ex$ cannot be in
equilibrium in a medium with such structure, which is why $E_z\not=0$
is needed in \eqref{eqn:E-Ez-bend} for bend distortions.  The proper,
full expression for $P_z$ in the bend case is given by
\begin{equation*}
  P_z^{\text{bend}} = \bigl( \chieperp n_x^2 + \chiepara n_z^2 \bigr) E_z +
  \Delta\chie n_x n_z E_0 ,
\end{equation*}
but we still have $\sigmaP=0$, since $P_z^{\text{bend}}=0$ at $z=0$
and $z=d$, due to the vanishing of both $n_x$ and $E_z$ on the
boundaries of the cell:
\begin{equation*}
  E_z = - \epsa E_0 \frac{n_x n_z}{\eperp n_x^2 + \epara n_z^2} ,
  \quad
  n_x(0) = n_x(d) = 0 .
\end{equation*}
The fact that $E_z\not=0$ in the case of bend distortions is entirely
due to the effective space charge from the induced polarization, which
is absent in the case of twist distortions.  Whether $E_z$ is positive
or negative, it leads to an increased magnitude of the local electric
field, $E^2=E_0^2+E_z^2>E_0^2$, and it emerges as soon as $\nhat$ is
perturbed from $\nhat_0$ (producing $n_x\not=0$ above).  It is also
the case that the additional electric-field component is in the $z$
direction, which is \emph{aligning} and weakens the dielectric torque
that is trying to rotate the director towards the horizontal.

The situation for electric-field \Freed\ transitions with $\epsa>0$
then is the following.  The splay transition is well analyzed in
\cite{gruler:scheffer:meier:72} and \cite{deuling:72}, and a good
textbook account is given in \cite[Sec.\,3.5]{stewart:04}.  The
instability threshold follows the magnetic-field analogy, and the
transition is second order for all relevant parameter values.  None of
this is in question here.  The twist transition is modeled by the free
energy \eqref{eqn:Fn-twist}, and as discussed above, the equilibrium
electric field remains \emph{uniform} for any twist distortions.  Thus
the analysis of the electric-field twist-\Freed\ transition is
completely aligned with that of the associated magnetic-field
transition.
% If one replaces the electric parameters in
% \eqref{eqn:Fn-twist} by their magnetic counterparts,
% \begin{equation*}
%   \eps_0 \mapsto \mu_0 , \quad \epara \mapsto 1+\chipara, \quad
%   \eperp \mapsto 1+\chiperp , \quad E_0 \mapsto H ,
% \end{equation*}
% then one obtains
% \begin{equation*}
%   \eps_0 \bigl( \epara n_x^2 + \eperp n_y^2 \bigr) E_0^2 ~ \mapsto ~
%   \mu_0 \Delta\chi H^2 n_x^2 + \mu_0(1+\chiperp) H^2 =
%   \chia (\bmH\cdot\nhat)^2 + \text{const} , \quad \bmH = H \ex .
% \end{equation*}
% The free energy that results is the same as would be obtained from
% $\calFH$ in \eqref{eqn:FH-FE} for this geometry.
One concludes that the local instability threshold is
\begin{equation*}
  \Ec = \frac{\pi}{d} \sqrt{\frac{K_2}{\eps_0\epsa}}
\end{equation*}
and that the transition is second order for any $K_2>0$ and
$\epara>\eperp>0$.  The magnetic-field analogy again holds.  The only
case that requires further analysis is the electric-field bend-\Freed\
transition, which we take up next.%  The claim in
%\cite{gruler:scheffer:meier:72} that the electric field is uniform for
%bend distortions seems to have contributed to some of the
%misunderstandings that found their way into \cite{pikin:91},
%\cite{degennes:prost:93}, and \cite{stewart:04}.

\section{Analysis of the electric-field bend-\Freed\ transition}

\label{sec:analysis-of-E-field-bend}

The electric-field bend-\Freed\ transition is modeled in terms of the
free energy \eqref{eqn:Fn-bend}, which can be analyzed in various
ways.  Here we utilize a \emph{perturbation expansion} to identify and
study the bifurcation at the point at which the uniform ground-state
solution of the Euler-Lagrange equations ceases to be locally stable.
This type of analytical approach has been widely used in the area of
liquid crystals by Schiller (see the review article \cite{schiller:90})
and others \cite{self:please:sluckin:02,napoli:06}.

We represent the director in terms of its tilt angle as
\begin{equation*}
  \nhat = \sin\theta\,\ex + \cos\theta\,\ez , \quad \theta = \theta(z) ,
\end{equation*}
for which the ground state corresponds to $\theta=0$.  In terms of
this representation, the free-energy functional \eqref{eqn:Fn-bend}
takes the form
\begin{equation*}%\label{eqn:Fth-bend}
  \calF[\theta] = \frac12 \int_0^d \Bigl[
  \bigl( K_1 \ssth + K_3 \ccth \bigr) \thz^2 -
  \eps_0 \frac{\eperp\epara}{\eperp\ssth+\epara\ccth} E_0^2 \Bigr]
  \td z .
\end{equation*}
We employ the following non-dimensionalization:
\begin{equation}\label{eqn:Fbar-thbar-bend}
\begin{gathered}
  \Fbar[\thbar] = \frac12 \int_0^{\pi} \Bigl[
  \bigl( 1 + \aK \ssthbar \bigr) \thbar_{\!,\zbar}^2 -
  \gamma \frac1{1 - \aeps \ssthbar} \Bigr] \td \zbar \\
  \Fbar = \frac{\calF}{K_3\pi/d} , ~~ \zbar = \frac{\pi z}{d} , ~~
  \thbar(\zbar) = \theta(z) \\
  \aK := \frac{K_1-K_3}{K_3} , ~~
  \aeps := \frac{\epara-\eperp}{\epara} , ~~
  \gamma := \frac{\eps_0\eperp E_0^2}{K_3\pi^2/d^2} .
\end{gathered}
\end{equation}
The dimensionless parameters $\aK$ and $\aeps$ represent the relative
anisotropies of the elastic constants and dielectric constants, while
the coupling constant $\gamma$ reflects the relative strength of the
electric field compared to distortional elasticity.

Equilibrium states are solutions of the associated Euler-Lagrange
equation, which (after dropping bars) is given by
% \begin{equation}\label{eqn:EL-theta-bar}
%   \frac{\td}{\td z} \bigl[ \bigl( 1 + \aK \ssth \bigr) \thz \bigr] +
%   \sth \cth \biggl[
%   \frac{\gamma\aeps}{\displaystyle(1-\aeps\ssth)^2} - \aK \thz^2 \biggr]
%   = 0 , ~~ \theta(0) = \theta(\pi) = 0 .
% \end{equation}
\begin{equation}\label{eqn:EL-theta-bar}
  \frac{\td}{\td z} \biggl[ \bigl( 1 + \aK \ssth \bigr)
  \frac{\td\theta}{\td z} \biggr] + \sth \cth \biggl[
  \frac{\gamma\aeps}{\displaystyle(1-\aeps\ssth)^2} -
  \aK \Bigl( \frac{\td\theta}{\td z} \Bigr)^{\!2} \biggr]
  = 0 , ~~ \theta(0) = \theta(\pi) = 0 .
\end{equation}
The ground state $\theta=0$ is a solution for any values of $\aK$,
$\aeps$, and $\gamma$.  We view $\gamma$ as the \emph{control
  parameter} and wish to determine for what value of $\gamma$ (given
$\aK$ and $\aeps$) the ground state becomes locally unstable and the
nature of the branch of nonuniform equilibrium solutions that
bifurcates from the ground-state branch at that point.  We parameterize
the bifurcating branch in terms of a parameter $\eta$ and assume the
validity of the expansions
\begin{equation}\label{eqn:bend-expansion}
\begin{aligned}
  \theta(z;\eta) &= \eta\,\theta_1(z) + \eta^3\,\theta_3(z) + \cdots \\
%  \theta(z;\eta) &= \eta\,\theta_1(z) + \eta^3\theta_3(z) + \cdots \\
  \gamma(\eta) &= \gamma_0 + \gamma_2\,\eta^2 + \gamma_4\,\eta^4 + \cdots
%  \gamma(\eta) &= \gamma_0 + \gamma_2\eta^2 + \gamma_4\eta^4 + \cdots
\end{aligned}
\end{equation}
for small $\eta$ (the ground state corresponding to $\eta=0$).  The
assumed form \eqref{eqn:bend-expansion} is based upon problem
symmetry: if the scalar field $\theta$ is a solution of
\eqref{eqn:EL-theta-bar}, then so is $-\theta$.  There is no intrinsic
scale for the parameter $\eta$, and we find it convenient to normalize
it via
\begin{equation}\label{eqn:bend-normalization}
  w_1 \Bigl( \frac{\td\gamma}{\td\eta} \Bigr)^{\!2} +
  w_2 \int_0^{\pi} \Bigl( \frac{\td\theta}{\td\eta} \Bigr)^{\!2} \td z =
  \text{const} ,
\end{equation}
using the values $w_1=6\pi\aeps^2$, $w_2=2$, and $\text{const}=\pi$
for convenience in what follows.  The parameter $\eta$ can be viewed
as a type of ``pseudo arc-length'' in $\gamma$-$\theta$ space.  At the
first three orders, the normalization condition gives rise to
constraints
\begin{equation*}%\label{eqn:bend-constraints}
\begin{aligned}
  O(1)\!: & ~~ w_2 \int_0^{\pi} \theta_1^2 \, \td z = \text{const} \\
  O(\eta)\!: & ~~ 0 = 0 \\
  O(\eta^2)\!: & ~~ 2 w_1 \gamma_2^2 +
                    3 w_2 \int_0^{\pi} \theta_1 \theta_3 \, \td z = 0 .
\end{aligned}
\end{equation*}

Substituting the expansions for $\theta$ and $\gamma$ into the
Euler-Lagrange equation and boundary conditions, we obtain at order
$O(\eta)$
% \begin{equation*}
%   \theta_1'' + \gamma_0 \aeps \theta_1 = 0 , ~~
%   \theta_1(0) = \theta_1(\pi) = 0 ,
% \end{equation*}
\begin{equation*}
  \frac{\td^2\theta_1}{\td z^2} + \gamma_0 \aeps \theta_1 = 0 , ~~
%  \frac{\td^2}{\td z^2} \theta_1 + \gamma_0 \aeps \theta_1 = 0 , ~~
  \theta_1(0) = \theta_1(\pi) = 0 ,
\end{equation*}
for which the first nontrivial bifurcating solution is given by
\begin{equation*}
  \gamma_0 \aeps = 1 , ~~ \theta_1 = c_{1,1} \sin z ,
\end{equation*}
with the constant $c_{1,1}$ to be determined.  Using this expression
for $\theta_1$ in the normalization condition at order $O(1)$ gives
\begin{equation*}
  w_2 c_{1,1}^2 \int_0^\pi \sin^2\!z \, \td z =
  w_2 c_{1,1}^2 \frac{\pi}2 = \text{const} .
\end{equation*}
With our choices $w_2=2$ and $\text{const}=\pi$, the above yields
$c_{1,1}=\pm1$.  Choosing $c_{1,1}=1$ determines $\theta_1=\sin z$.

The problem for $\theta_3$ arises at order $O(\eta^3)$:
% \begin{equation*}
%   \theta_3'' + \theta_3 + a_{3,1} \sin z + a_{3,3} \sin 3z = 0 , ~~
%   \theta_3(0) = \theta_3(\pi) = 0 ,
% \end{equation*}
\begin{equation*}
  \frac{\td^2\theta_3}{\td z^2} + \theta_3 + a_{3,1} \sin z +
%  \frac{\td^2}{\td z^2} \theta_3 + \theta_3 + a_{3,1} \sin z +
  a_{3,3} \sin 3z = 0 , ~~ \theta_3(0) = \theta_3(\pi) = 0 ,
\end{equation*}
with
\begin{equation*}
  a_{3,1} = \frac12 [ ( 3 + 2 \gamma_2 ) \aeps - \aK - 1 ] , ~~
  a_{3,3} = \frac16 ( 3 \aK - 3 \aeps + 1 ) .
\end{equation*}
Solvability requires that $a_{3,1}=0$ (see \cite[Ch.\,15]{nayfeh:81}),
giving
\begin{equation*}
  \gamma_2 \aeps = \frac12 ( \aK - 3 \aeps + 1 ) .
\end{equation*}
The solution $\theta_3$ then becomes
\begin{equation}\label{eqn:bend-c33}
  \theta_3 = c_{3,1} \sin z + c_{3,3} \sin 3z , ~~
  c_{3,3} = \frac1{48} ( 3 \aK - 3 \aeps + 1 ) ,
\end{equation}
with $c_{3,1}$ to be determined.  Substituting the expressions for
$\gamma_2$, $\theta_1$, and $\theta_3$ into the normalization
condition at order $O(\eta^2)$ produces
\begin{equation*}
  w_1 \Bigl( \frac{\aK-3\aeps+1}{\aeps} \Bigr)^{\!2} +
  3 \pi w_2 c_{3,1} = 0 ,
\end{equation*}
which gives
\begin{equation}\label{eqn:bend-c31}
  c_{3,1} = - ( \aK - 3 \aeps + 1 )^2
\end{equation}
using our chosen weights $w_1=6\pi\aeps^2$ and $w_2=2$.

At this stage, we have all the information that we require.  In terms
of our dimensionless variables, we have
\begin{align*}
  \theta &= \eta \sin z + \eta^3 ( c_{3,1} \sin z + c_{3,3} \sin 3z ) +
            \cdots \\
  \aeps \gamma &= 1 + \frac12 ( \aK - 3 \aeps + 1 ) \eta^2 + \cdots ,
\end{align*}
with $c_{3,1}$ and $c_{3,3}$ as given in \eqref{eqn:bend-c31} and
\eqref{eqn:bend-c33}.  Our choice of weights in
\eqref{eqn:bend-normalization} has led to a normalization in which the
parameter $\eta$ corresponds to the \emph{amplitude} of the leading
instability mode, the longest-wavelength mode $\sin\zbar$.  In terms
of the original variables, the expansions read
\begin{align*}
  \theta &= \eta \sin\frac{\pi z}{d} + \eta^3 \biggl\{ - \Bigl(
            \frac{K_1}{K_3} - 3 \frac{\epsa}{\epara} \Bigr)^{\!2}
            \sin \frac{\pi z}{d} \\
         &\mathrel{\phantom{=}} {} + \frac1{48} \Bigl[ 3 \Bigl( \frac{K_1}{K_3} -
          \frac{\epsa}{\epara} \Bigr) - 2 \Bigr]
          \sin \frac{3\pi z}{d} \biggr\} + \cdots \\
  E_0^2 &= \frac{\epara}{\,\eperp} \frac{\pi^2}{d^2} \frac{K_3}{\eps_0\epsa}
           \Bigl[ 1 + \frac12 \Bigl( \frac{K_1}{K_3} - 3
           \frac{\epsa}{\epara} \Bigr) \eta^2 + \cdots \Bigr] .
\end{align*}
We see that the critical threshold here is \emph{elevated} from that
predicted by the magnetic-field analogy:
%and \cite[Eqn.\,(3.208)]{stewart:04}:
\begin{equation*}
  \Ec = \frac{\pi}{d} \sqrt{\frac{\epara}{\,\eperp}}
  \sqrt{\frac{K_3}{\eps_0\epsa}} , ~ \text{with} ~
  \frac{\epara}{\,\eperp} > 1 ~~ \text{vs} ~~
  \Ec = \frac{\pi}{d} \sqrt{\frac{K_3}{\eps_0\epsa}} ~ \leftrightarrow ~
  \Hc = \frac{\pi}{d} \sqrt{\frac{K_3}{\chia}} .
\end{equation*}
The sign of the coefficient in front of $\eta^2$ in the expansion for
$E_0^2$ above indicates whether the bifurcation is
\emph{super-critical} (positive coefficient) or \emph{sub-critical}
(negative coefficient).  The former case ($K_1/K_3-3\,\epsa/\epara>0$)
is consistent with a \emph{second-order} transition, while the latter
would be associated with a transition of \emph{first order}.

The formulas above for the critical threshold and the crossover from
first order to second order agree with predictions in
\cite{arakelyan:karayan:chilingaryan:84}, where they were deduced in a
different way.  The question of when a \Freed\ transition can be first
order has been well studied elsewhere---see \cite{frisken:palffy:89c}
and references contained therein.  We note that the elevation of the
threshold here is \emph{not small}: using data from
\cite[Table\,D.3]{stewart:04} for the liquid-crystal material 5CB near
$26^{\circ}$C, we have
\begin{equation*}
  \epara = 18.5, ~ \eperp = 7 ~ \Rightarrow ~
  \sqrt{\frac{\epara}{\,\eperp}} \doteq 1.63 ,
\end{equation*}
a 63\% increase.  For the same material and temperature (with data
from the same source), we have $K_1=6.2\times10^{-12}$\,J/m and
$K_3=8.2\times10^{-12}$\,J/m.  With these material parameters, the
quantity $K_1/K_3-3\,\epsa/\epara = - 1.1 < 0$\,; so the transition
should be \emph{first order}, as was found in experiments reported in
\cite{frisken:palffy:89,frisken:palffy:89b}.

\begin{figure}
  \centering
  \includegraphics[width=.32\textwidth]{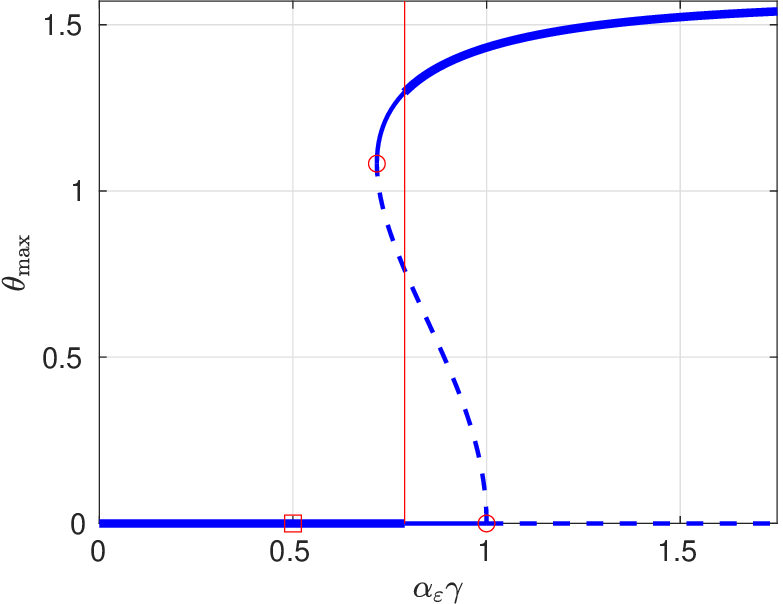}
  \includegraphics[width=.32\textwidth]{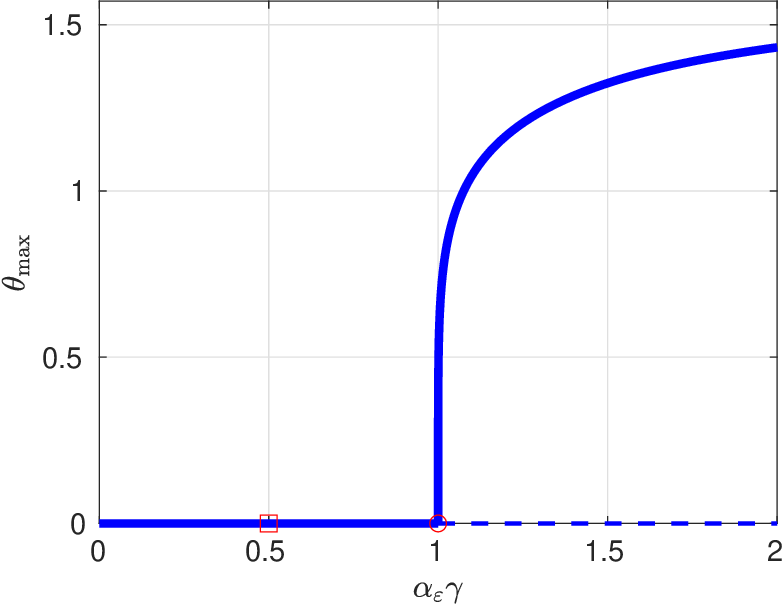}
  \includegraphics[width=.32\textwidth]{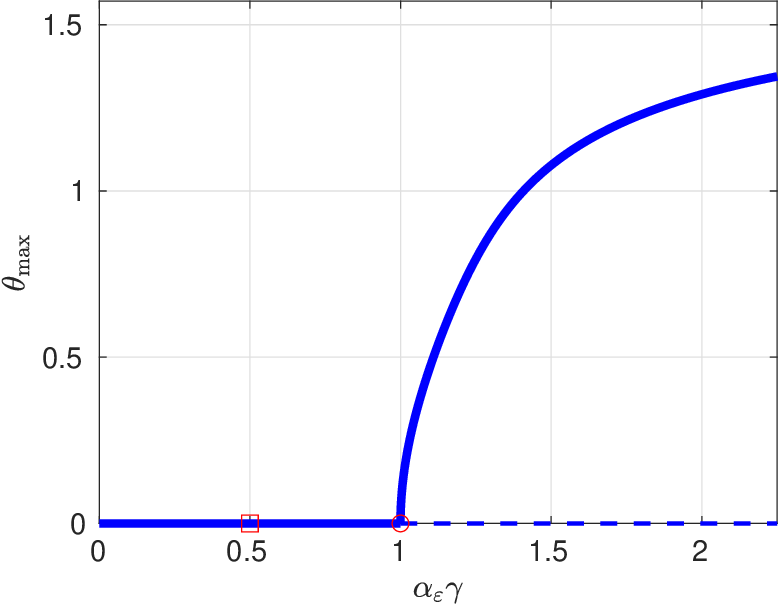}
  \caption{Upper halves of bifurcation diagrams for electric-field
    bend-\Freed\ transitions in terms of reduced units as defined in
    \eqref{eqn:Fbar-thbar-bend}.  Maximum tilt angle $\theta_{\max}$
    versus reduced electric field strength $\aeps\gamma$.  Dashed line
    (locally unstable), solid line (locally stable, metastable), heavy
    solid line (globally stable, minimum free energy).  For all three
    figures, the relative dielectric anisotropy is fixed at
    $\aeps=0.5$.  The relative anisotropy of the elastic constants
    varies as follows: $\aK=-0.5$ (left, first-order transition), 0.5
    (center, crossover), 1.5 (right, second-order transition).  The
    boxes on the horizontal axes are at the location of the
    instability threshold predicted by the magnetic-field analogy
    (treating the electric field as uniform), given by
    $\aeps\gamma=1-\aeps=0.5$ in these units.}
  \label{fig:bif-diagrams}
\end{figure}

In Figure\,\ref{fig:bif-diagrams}, the upper halves of bifurcation
diagrams are plotted for three sets of dimensionless parameters based
upon numerical modeling of \eqref{eqn:EL-theta-bar} using a numerical
bifurcation package, as described in \cite[Secs.\,6.2 and
6.3]{richards:06}.  The parameter values were chosen to illustrate the
crossover from a first-order transition to a second-order transition,
which occurs at $\aK=0.5$ when $\aeps=0.5$ (as in
Figure\,\ref{fig:bif-diagrams}).  The phenomena of an elevated local
instability threshold and a first-order transition can be seen as
related: the elevation of the threshold necessitates a sub-critical
bifurcation in order for the branch to reach distorted equilibrium
configurations of the director field that are locally stable for
electric-field strengths of lesser amounts (for certain combinations
of material constants).

The natural question at this point is why the local-instability
threshold is elevated for the electric-field bend transition but not
for the splay or twist transitions.  A partial answer has already been
provided by the observation that the electric field remains
\emph{uniform} for \emph{twist} distortions but develops
\emph{inhomogeneity} with \emph{bend} distortions.  This does not
explain, however, why this should lead to an elevation of the
instability threshold---in the splay geometry, inhomogeneity of the
electric field develops as well, but that does not lead to an
elevation of the threshold.  An answer is provided by a careful
development of a criterion for local stability of a coupled director
field and electric field, generalizing \eqref{eqn:FH-neps} and
\eqref{eqn:H-local-stability}.  This was done in a general setting in
\cite{gartland:21} and is discussed next.

\section{Electric-field-induced instabilities}

\label{sec:E-field-induced-instabilities}

Whether the coupling between the electric field and the director field
will affect the threshold of local instability depends on
\emph{dielectric polarization}, in particular on how the induced
polarization changes at a bifurcation from a branch of ground-state
director configurations.  We illustrate this by presenting a necessary
condition for local stability and showing how it correctly predicts
the behaviors that we have observed with the three basic
electric-field \Freed\ transitions.  Additional examples are given
that evidence the broader applicability of the development, including
systems involving twisted nematics, planar cholesterics, smectic-A
phases, and nematic films that exhibit periodic instabilities.

\subsection{A necessary condition for local stability}

As modeled by the free-energy functional $\calFE$ in
\eqref{eqn:FH-FE}, a locally stable director field $\nhat$ will be a
local minimizer of $\calFE[\nhat,\varphi]$ subject to the constraints
\begin{equation*}
  |\nhat| = 1 ~ \text{ and } ~
  \div \bigl[ \epstensor(\nhat)\nabla\varphi \bigr] = 0 , ~
  \text{ on } \Omega ,
\end{equation*}
plus appropriate boundary conditions on both $\nhat$ and $\varphi$.
As such, a sufficiently smooth $\nhat$ must necessarily satisfy the
Euler-Lagrange equations
\begin{equation}\label{eqn:E-Euler-Lagrange}
  - \div \Bigl( \frac{\upartial\We}{\upartial\nabla\nhat} \Bigr) +
  \frac{\upartial\We}{\upartial\nhat} = \lambda \nhat +
  \eps_0 \epsa (\nabla\varphi\cdot\nhat) \nabla\varphi , ~
  \text{ on } \Omega .
\end{equation}
In order to determine if an equilibrium pair $(\nhat_0,\varphi_0)$ is
locally minimizing, we require an expression for
$\calFE[\nhat,\varphi]$ for admissible pairs $(\nhat,\varphi)$ close
to $(\nhat_0,\varphi_0)$, generalizing \eqref{eqn:FH-neps}.

A perturbation of the director field
$\nhat_0\mapsto\nhat_0+\epsilon\bmu$ will satisfy all the constraints
on $\nhat$ to first order in $\epsilon$ if $\bmu$ is \emph{transverse}
to $\nhat_0$ ($\nhat_0\cdot\bmu=0$ on $\Omega$) and satisfies
\emph{homogeneous} boundary conditions on $\partial\Omega$.  We need
to know under what conditions a perturbation
$(\nhat_0,\varphi_0)\mapsto(\nhat_0+\epsilon\bmu,\varphi_0+\epsilon\psi)$
will satisfy to first order the constraint on $\varphi$ as well
(namely $\div[\epstensor(\nhat)\nabla\varphi]=0$).  To deduce this, we
use \eqref{eqn:eps-tensor} to expand
\begin{equation*}
  \epstensor(\nhat_0+\epsilon\bmu) \nabla(\varphi_0+\epsilon\psi) =
  \epstensor(\nhat_0) \nabla\varphi_0 + \epsilon \bigl[
  \Delta\chie(\nhat_0\otimes\bmu+\bmu\otimes\nhat_0) \nabla\varphi_0 +
  \epstensor(\nhat_0) \nabla\psi \bigr] + O(\epsilon^2) .
\end{equation*}
Here we have also used $\eps_0\epsa=\Delta\chie$ from
\eqref{eqn:eps-vs-chi} to express the above in terms of electric
susceptibilities.  Since $\div[\epstensor(\nhat_0)\nabla\varphi_0]=0$
by assumption, in order for the constraint
$\div[\epstensor(\nhat)\nabla\varphi]=0$ to be preserved at first
order, $\psi$ must satisfy
\begin{subequations}
  \begin{equation}\label{eqn:div-grad-psi}
    \div[\epstensor(\nhat_0)\nabla\psi] = \div\bmp_0 ,
    ~ \text{ on } \Omega ,
  \end{equation}
where
  \begin{equation}\label{eqn:pzero-Ezero}
    \bmp_0 := \Delta\chie (\nhat_0\otimes\bmu+\bmu\otimes\nhat_0) \bmE_0 ,
    \quad \bmE_0 = - \nabla\varphi_0 .
  \end{equation}
\end{subequations}
As is the case with the director field and its perturbations, the
field $\psi$ must be such that $\varphi_0+\epsilon\psi$ satisfies the
boundary conditions that $\varphi$ must satisfy.  While more general
conditions are considered in \cite{gartland:21}, here we assume, for
simplicity, that the electric potential $\varphi$ is fixed on the
entire boundary $\partial\Omega$, which implies that all such
perturbative fields $\psi$ must \emph{vanish} on $\partial\Omega$.
Thus, given an equilibrium pair $(\nhat_0,\varphi_0)$ and director
perturbation $\nhat_0\mapsto\nhat_0+\epsilon\bmu$, the associated
perturbation of the electric potential
$\varphi_0\mapsto\varphi_0+\epsilon\psi$ must satisfy an electrostatic
potential problem consisting of \eqref{eqn:div-grad-psi} plus
homogeneous boundary conditions on $\partial\Omega$.

The vector field $\bmp_0$ represents the \emph{first-order} change in
induced polarization associated with the perturbation
$\nhat_0\mapsto\nhat_0+\epsilon\bmu$, and $-\div\bmp_0$ acts as an
effective \emph{volume charge density} driving $\psi$ and the
associated perturbation of the electric field
$\bmE_0\mapsto\bmE_0-\epsilon\nabla\psi$.  The only solution of
$\div[\epstensor(\nhat_0)\nabla\psi]=0$ satisfying $\psi=0$ on
$\partial\Omega$ is the \emph{zero solution}, and we conclude that
% $\psi=0$ on $\Omega$ \emph{if and only if} $\div\bmp_0=0$ on $\Omega$.
% \begin{equation*}
%   \psi=0 \text{ on } \Omega ~ \Leftrightarrow ~
%   \div\bmp_0 = 0 \text{ on } \Omega .
% \end{equation*}
% \begin{equation*}
%   \psi=0 \text{ on } \Omega ~\, \text{if and only if} ~
%   \div\bmp_0 = 0 \text{ on } \Omega .
% \end{equation*}
\begin{equation*}
  \psi=0 \text{ on } \Omega ~ \text{\it if and only if} ~
  \div\bmp_0 = 0 \text{ on } \Omega .
\end{equation*}
Thus if $\div\bmp_0=0$ on $\Omega$, then the constraint
$\div[\epstensor(\nhat)\nabla\varphi]=0$ will be preserved to first
order under the perturbation $\nhat_0\mapsto\nhat_0+\epsilon\bmu$ with
an \emph{unperturbed} electric field $\bmE_0=-\nabla\varphi_0$.
Changes in the director field cause changes in the electric field in
general, but in the case $\div\bmp_0=0$, these changes come in at
\emph{higher order}.  Making use of the information gleaned thus far,
the following is established in \cite{gartland:21}.

Let $(\nhat_0,\varphi_0)$ be an equilibrium pair associated with
$\calFE$ in \eqref{eqn:FH-FE} as above, and let $(\neps,\phieps)$ be
an admissible pair perturbed from $(\nhat_0,\varphi_0)$ constructed in
the form
\begin{equation*}
  \neps = \frac{\nhat_0+\epsilon\bmv}{|\nhat_0+\epsilon\bmv|} , \quad
  \div \bigl[ \epstensor(\neps)\nabla\phieps \bigr] = 0 ,
\end{equation*}
where $\bmv=\bmzero$ on $\partial\Omega$ and the potential $\phieps$
satisfies the boundary conditions that $\varphi$ must satisfy.  The
free energy $\calFE[\neps,\phieps]$ admits the following expansion for
small $\epsilon$:
\begin{equation*}
  \calFE[\neps,\phieps] = \calFE[\nhat_0,\varphi_0] +
  \frac12 \epsilon^2 \Bigl[
  \udelta^2_{\nhat\nhat} \calFE[\nhat_0,\varphi_0](\bmu) +
  \int_\Omega \epstensor(\nhat_0) \nabla\psi \cdot \nabla\psi \, \td V -
  \int_\Omega \lambda_0 |\bmu|^2 \, \td V \Bigr] + o(\epsilon^2) ,
\end{equation*}
where $\bmu=\bfP(\nhat_0)\bmv$ as in \eqref{eqn:neps-u-P-v} and $\psi$
solves the problem \eqref{eqn:div-grad-psi} with $\psi=0$ on
$\partial\Omega$.  Here $\lambda_0$ is the Lagrange multiplier field
associated with the solution $\nhat_0$ of
\eqref{eqn:E-Euler-Lagrange}, and $\udelta^2_{\nhat\nhat}\calFE$
denotes the second variation of $\calFE$ with respect to $\nhat$:
\begin{multline*}
  \udelta^2_{\nhat\nhat} \calFE[\nhat,\varphi](\bmv) = \int_\Omega \Bigl(
  \frac{\upartial^2 W}{\upartial n_i \upartial n_k} v_i v_k +
  2 \frac{\upartial^2 W}{\upartial n_i \upartial n_{k,l}} v_i v_{k,l} +
  \frac{\upartial^2 W}{\upartial n_{i,j} \upartial n_{k,l}} v_{i,j} v_{k,l}
  \Bigr) \td V , \\
  W = \We - \frac12 \epstensor(\nhat) \nabla\varphi \cdot \nabla\varphi .
\end{multline*}
We conclude that in order for an equilibrium director field $\nhat_0$
with associated Lagrange multiplier field $\lambda_0$ and coupled
electric potential field $\varphi_0$ to be \emph{locally stable}, it
must necessarily satisfy
\begin{equation}\label{eqn:E-local-stability}
  \udelta^2_{\nhat\nhat} \calFE[\nhat_0,\varphi_0](\bmu) +
  \int_\Omega \epstensor(\nhat_0) \nabla\psi \cdot \nabla\psi \, \td V -
  \int_\Omega \lambda_0 |\bmu|^2 \, \td V \ge 0
\end{equation}
for all test fields $\bmu$ that are transverse to $\nhat_0$
($\nhat_0\cdot\bmu=0$ on $\Omega$) and satisfy $\bmu=\bmzero$ on
$\partial\Omega$, with the scalar field $\psi$ (which depends on
$\bmu$) the solution of \eqref{eqn:div-grad-psi} subject to $\psi=0$
on $\partial\Omega$.

The necessary condition above generalizes
\eqref{eqn:H-local-stability}.  The first and third terms in
\eqref{eqn:E-local-stability} are already present in
\eqref{eqn:H-local-stability}; while the second term is new.  The
integral involving $\lambda_0$ is again associated with the pointwise
constraint $|\nhat|=1$.  The second term is associated with the
constraint on the electric potential
($\div[\epstensor(\nhat)\nabla\varphi]=0$) and compensates for the
fact that the perturbation $\psi$ preserves this only to first order.
The perturbation of the electric potential, $\psi$, is slaved to the
perturbation of the director field, $\bmu$, just as the electric
potential $\varphi$ is slaved to the director field $\nhat$.  The
dielectric tensor $\epstensor(\nhat_0)$ is \emph{positive definite};
so the middle term in \eqref{eqn:E-local-stability} is \emph{strictly
  positive} unless $\nabla\psi$ is \emph{identically zero} on
$\Omega$.  The presence of a positive second term in
\eqref{eqn:E-local-stability} increases the energy barrier that must
be overcome to destabilize the ground state, and it is this term that
causes the increase in the local-instability threshold.  That this
middle term is \emph{nonnegative} emerges in a natural way in the
calculus of deriving \eqref{eqn:E-local-stability} (see
\cite{gartland:21}); the nonnegativity can be viewed as a consequence
of the minimax nature of the problem.  A perturbation of a ground
state director field will cause a change in the induced polarization
in general; the discriminating factor is whether this leads to a
\emph{first-order} or \emph{higher-order} effect.

In an electric-field transition, then, there are \emph{two}
stabilizing influences: distortional elasticity (which penalizes
nonuniformity of the director field) and electrostatic energy (which
penalizes spontaneous additions to the electric field).  The former is
present in magnetic-field transitions; the latter is not.  The
necessary condition for local stability \eqref{eqn:E-local-stability}
provides us with two useful pieces of information.  First, it shows us
that the coupling between the electric field and the director field
can influence an instability threshold only if there exists a director
perturbation $\bmu$ for which $\div\bmp_0$ is \emph{not} identically
zero on $\Omega$, since this is the only way to obtain a $\psi$ such
that $\nabla\psi$ is not identically zero on $\Omega$, and second, it
shows us that the effect can only be to \emph{elevate} an instability
threshold, \emph{never} to \emph{lower} it (since the middle term in
\eqref{eqn:E-local-stability} can only be \emph{positive} or \emph{zero}).
% While \eqref{eqn:E-local-stability} is a little ``unwieldy'' for
% deriving formulas for instability thresholds, the criterion
% involving $\div\bmp_0$ provides a simple test for when the
% coupling between the director field and the electric field can cause
% an elevated threshold, which we now demonstrate.
The criterion involving $\div\bmp_0$ provides a simple test for when
the coupling between the director field and the electric field can
cause an elevated threshold, which we now demonstrate.

\subsection{Application to classical electric-field \Freed\ transitions}

\label{sec:application-to-classical-Freed}

% Our test for electric-field-induced transitions involves the
% first-order change in induced polarization $\bmp_0$ in
% \eqref{eqn:pzero-Ezero} and is the following.  If $\div\bmp_0=0$ on
% $\Omega$ for \emph{all} admissible variations $\bmu$, then the
% instability threshold will \emph{not} be elevated and will conform to
% the magnetic-field analogy.  If, however, there exists an admissible
% variation $\bmu$ such that $\div\bmp_0$ is \emph{not} identically zero
% on $\Omega$, then the instability threshold \emph{will} be elevated.
% Consider applying this test to the electric-field splay, twist, and
% bend \Freed\ geometries (with $\epsa>0$).

Our test for electric-field-induced transitions involves the
first-order change in induced polarization $\bmp_0$ in
\eqref{eqn:pzero-Ezero} and is the following.
\begin{quote}
  If $\div\bmp_0=0$ on $\Omega$ for \emph{all} admissible director
  variations $\bmu$, then the instability threshold will \emph{not} be
  elevated and will conform to the magnetic-field analogy.  If,
  however, there exists an admissible variation $\bmu$ such that
  $\div\bmp_0$ is \emph{not} identically zero on $\Omega$, then the
  instability threshold \emph{will} be elevated.
\end{quote}
We show how this test can be applied to the electric-field splay,
twist, and bend \Freed\ transitions (with $\epsa>0$).

The splay geometry is illustrated in Figure\,\ref{fig:chia-pos-geoms}
left, with the director field confined to the $x$-$z$ tilt plane
($\nhat=n_x(z)\ex+n_z(z)\ez$).  The ground-state director field
$\nhat_0$, ground-state electric field $\bmE_0$, and admissible
director variations $\bmu$ are given by
\begin{equation*}
  \nhat_0 = \ex , \quad \bmE_0 = E_0 \ez , \quad \bmu = w(z) \ez ,
\end{equation*}
where $w$ is a sufficiently smooth function satisfying $w(0)=w(d)=0$.
In terms of these, $\bmp_0$ takes the form
\begin{equation*}
  \bmp_0 = \Delta\chie (\nhat_0\otimes\bmu+\bmu\otimes\nhat_0) \bmE_0 =
  \Delta\chie \bigl[ (\bmu\cdot\bmE_0)\nhat_0 +
  (\nhat_0\cdot\bmE_0)\bmu \bigr] = \Delta\chie E_0 w(z) \ex ,
\end{equation*}
and it follows that
\begin{equation*}
  \div \bmp_0 = \Delta\chie E_0 \frac{\upartial}{\upartial x} w(z) = 0
\end{equation*}
for all such perturbations $\bmu$.  We conclude that the instability
threshold should \emph{not} be elevated, which is consistent with the
analyses in \cite{gruler:scheffer:meier:72}, \cite{deuling:72}, and
\cite[Sec.\,3.5]{stewart:04}.  The twist geometry
(Figure\,\ref{fig:chia-pos-geoms} center) yields a similar result:
\begin{gather*}
  \nhat_0 = \ey , \quad \bmE_0 = E_0 \ex , \quad \bmu = v(z) \ex \\
  \bmp_0 = \Delta\chie E_0 v(z) \ey ~ \Rightarrow ~
  \div \bmp_0 = \Delta\chie E_0 \frac{\upartial}{\upartial y} v(z) = 0 .
\end{gather*}
Again, no elevation of the instability threshold is expected,
consistent with our analysis in
Section~\ref{sec:models-of-E-field-transitions}.

Only in the case of the bend geometry (Figure\,\ref{fig:chia-pos-geoms}
right) do we encounter a departure from the script, for in that case,
\begin{gather*}
  \nhat_0 = \ez , \quad \bmE_0 = E_0 \ex , \quad \bmu = u(z) \ex \\
  \bmp_0 = \Delta\chie E_0 u(z) \ez ~ \Rightarrow ~
  \div \bmp_0 = \Delta\chie E_0 \frac{\upartial}{\upartial z} u(z) \not= 0 .
\end{gather*}
In this case, the first-order change in the induced polarization is
\emph{not} divergence free for all admissible $\bmu$, and there will
be corresponding \emph{positive} middle terms in
\eqref{eqn:E-local-stability}, which will lead to an elevated
instability threshold (as predicted in
\cite{arakelyan:karayan:chilingaryan:84} and confirmed by our analysis
in Section~\ref{sec:analysis-of-E-field-bend}).  It is possible to use
\eqref{eqn:E-local-stability} to deduce the formula for the elevated
threshold,
\begin{equation*}
  \Ec^2 = \frac{\epara}{\,\eperp} \frac{\pi^2}{d^2}
  \frac{K_3}{\eps_0\epsa} .
\end{equation*}
See \cite[Sec.\,4.2]{gartland:21}.

In both the splay geometry and the bend geometry, inhomogeneity of the
electric field is brought about by the \Freed\ transition.  The
difference between the two is that in the splay geometry, this is a
\emph{second-order} effect, while in the bend geometry, it is
\emph{first order}.  This can be seen by a closer examination of the
changes in polarization that are caused by the changes in the director
field at the onset of the instability, which we now show.

In the splay geometry, at the instability onset (with the electric
field treated as uniform), we have
\begin{equation*}
  \bmE = E_0 \, \ez , \quad
  \nhat = \cos\theta \, \ex + \sin\theta \, \ez , \quad
  \theta = \theta(z) ,
\end{equation*}
where $\theta$ is \emph{small} and satisfies $\theta(0)=\theta(d)=0$,
the ground state $\nhat_0=\ex$ corresponding to $\theta=0$.  From
\eqref{eqn:P-chie}, the polarization is given by
\begin{equation*}
  \bmP = \chieperp E_0 \, \ez + \Delta\chie E_0 \sin\theta
  ( \cos\theta \, \ex + \sin\theta \, \ez ) ,
\end{equation*}
so that $\bmP_0=\chieperp E_0\,\ez$ in the ground state, and the change
at onset (for small $\theta$) is
\begin{equation*}
  \bmP - \bmP_0 = \Delta\chie E_0 \bigl[
  \theta(z) \ex + \theta(z)^2 \ez + O(\theta^3) \bigr] .
\end{equation*}
Thus there is a first-order change in the $\ex$ component of
$\bmP-\bmP_0$. All fields are assumed to be \emph{uniform} in $x$ and
$y$, however; so this does not affect $\div\bmP$.  The change in the
$\ez$ component is \emph{second order} in $\theta$, and as a
consequence, \emph{no change} in the electric field is needed to
preserve $\div\bmD=0$ to first order---a second-order change in $\bmE$
suffices to maintain $\div\bmD=0$ at second order.

In the bend geometry, on the other hand, we have
\begin{equation*}
  \bmE = E_0 \, \ex , \quad \nhat = \sin\theta \, \ex + \cos\theta \, \ez ,
\end{equation*}
giving
\begin{equation*}
  \bmP = \chieperp E_0 \, \ex + \Delta\chie E_0 \sin\theta
  ( \sin\theta \, \ex + \cos\theta \, \ez )
\end{equation*}
and
\begin{equation*}
  \bmP - \bmP_0 = \Delta\chie E_0 \bigl[
  \theta(z) \ez + \theta(z)^2 \ex + O(\theta^3) \bigr] .
\end{equation*}
The change in the $\ez$ component is now \emph{first order} in
$\theta$ (the same as $\bmp_0=\Delta\chie E_0 u(z) \ez$ previously),
and a first-order change in $\bmE$ is required to maintain
$\div\bmD=0$.

Closely related to the electric-field bend-\Freed\ transition with
$\epsa>0$ above is the splay transition with $\epsa<0$
(Figure\,\ref{fig:chia-neg-geoms} left), for which
\begin{equation*}
  \bmE = E_0 \, \ex , \quad \nhat = \cos\theta \, \ex + \sin\theta \, \ez .
\end{equation*}
For this we have
\begin{equation*}
  \bmP = \chieperp E_0 \, \ex + \Delta\chie E_0 \cos\theta
  ( \cos\theta \, \ex + \sin\theta \, \ez ) ,
\end{equation*}
which gives
\begin{equation*}
  \bmP - \bmP_0 = \Delta\chie E_0 \bigl[ (\cos^2\theta-1) \, \ex +
  \sin\theta \cos\theta \, \ez \bigr] = \Delta\chie E_0 \bigl[
  - \theta(z)^2 \ex + \theta(z) \ez + O(\theta^3) \bigr] ,
\end{equation*}
and the first-order change in the $\ez$ component again necessitates a
first-order correction in $\bmE$ in order to maintain $\div\bmD=0$.
The distinction rests entirely on the relative orientations of the
ground-state director field $\nhat_0$, the ground-state electric field
$\bmE_0$, and the admissible variations of the director field $\bmu$
(and the spatial dependencies of these).

\subsection{Applications to generalized \Freed\ transitions}

In addition to the classical \Freed\ transitions, there are other
examples of field-induced instabilities in liquid-crystal systems,
which are sometimes referred to as ``generalized \Freed\
transitions,'' and the necessary condition for local stability
\eqref{eqn:E-local-stability} applies to them as well.  It is also the
case that the coupling between a director field and an electric field
is essentially the same in any liquid-crystal phase; so while the
criterion \eqref{eqn:E-local-stability} and related ideas were
developed in \cite{gartland:21} in the context of a \emph{nematic}
liquid crystal, it is expected that they remain valid for
\emph{cholesteric} and \emph{smectic} liquid crystals, for example.
In this section, we consider some cases from these broader areas.

\subsubsection{Twisted nematics and cholesterics}

\label{sec:twisted-nematics-and-cholesterics}

\begin{figure}
  \centering
  \hfill
  \includegraphics[width=.32\textwidth]{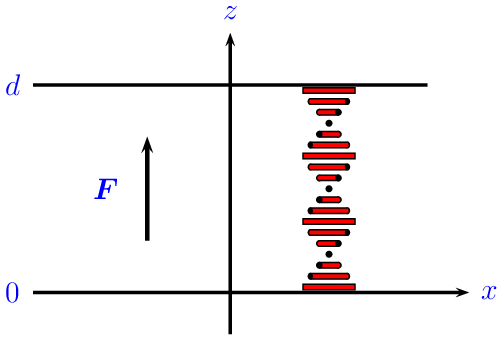}
  \hfill
  \includegraphics[width=.32\textwidth]{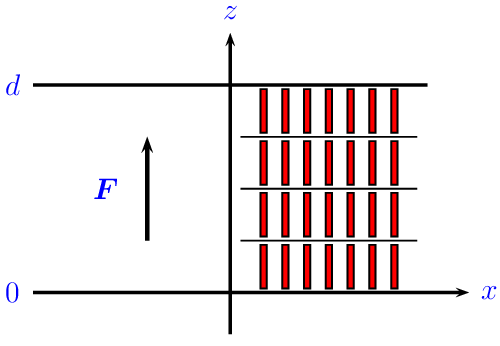}
  \hspace*{\fill}
  \caption{Other systems with instabilities induced by \emph{magnetic}
    or \emph{electric} fields ($\bmF=\bmH$ or $\bmF=\bmE$):
    \emph{twisted nematic} or \emph{cholesteric} (left, $\chia>0$ or
    $\epsa>0$), \emph{smectic\,A} (right, $\chia<0$ or $\epsa<0$).}
  \label{fig:cholesteric-smectic-geoms}
\end{figure}

Planar twisted configurations of the form in
Figure\,\ref{fig:cholesteric-smectic-geoms} left can be realized with
nematic liquid crystals under the influence of boundary conditions or
with cholesteric materials, for which such twisting states are natural
even in the absence of any boundary influence.  Ground states for
director fields of the type pictured have the form
\begin{equation*}
  \nhat_0 = \cos q_0 z \, \ex + \sin q_0 z \, \ey .
\end{equation*}
It is known that for modest values of $q_0d$, materials with positive
anisotropy will, at a critical field strength, undergo a transition of
\Freed\ type to a distorted configuration that remains uniform in $x$
and $y$ but has a non-vanishing $z$ component of the director.  This
basic, well-studied instability underlies the functioning of the
``Twisted Nematic Cell'' (TNC) and ``Supertwisted Nematic'' (STN)
device, for which $q_0d=\pi/2$ and $3\pi/2$ respectively
\cite[Sec.\,3.7]{stewart:04}.  The necessary condition for local
stability in \eqref{eqn:E-local-stability} and associated test in
Section~\ref{sec:application-to-classical-Freed} correctly indicate
that there should be \emph{no elevation} of the local instability
threshold in such systems when using an electric field instead of
magnetic field (with the magnetic parameters replaced by appropriate
electric parameters in the threshold formula).  This can be seen as
follows.

The ground-state electric field is $\bmE_0=E_0\ez$, and admissible
variations are of the form
\begin{equation*}
  \bmu = u(z) \ex + v(z) \ey + w(z) \ez , ~~ \bmu(0) = \bmu(d) = \bmzero .
\end{equation*}
The requirement $\nhat_0\cdot\bmu=0$ imposes a constraint on $u$ and $v$,
%The constraint $\nhat_0\cdot\bmu=0$ imposes
% \begin{equation*}
%   u(z) \cos q_0 z + v(z) \sin q_0 z = 0 , ~~ 0 < z < d ,
% \end{equation*}
but this does not affect the analysis.  In terms of these fields, the
first-order change in induced polarization \eqref{eqn:pzero-Ezero} is
given by
\begin{equation*}
  \bmp_0 = \Delta\chie \bigl[ (\bmu\cdot\bmE_0) \nhat_0 +
  (\nhat_0\cdot\bmE_0) \bmu \bigr] =
  \Delta\chie E_0 w(z) ( \cos q_0 z \ex + \sin q_0 z \ey ) ,
\end{equation*}
which implies that $\div\bmp_0=0$ for all such admissible
perturbations, and we conclude that the magnetic approximation should
apply.  The switching threshold for such a twist configuration in a
magnetic field was derived in \cite{leslie:70} and is recounted in
\cite[Sec.\,3.7]{stewart:04}.  It was assumed in
\cite{schadt:helfrich:71} and other papers that the same formula but
with electric parameters would give the threshold in an electric
field.  The validity of this magnetic-field analogy was established in
subsequent analyses, including \cite{fraser:78,raynes:86,schiller:89}.

It is known that for larger values of $q_0d$ the transition discussed
above is preceded (at a lower field strength) by an instability to a
configuration that is periodic or doubly periodic in the $x$-$y$
plane---a contemporary overview of this and related issues can be
found in \cite{blanc:durey:kamien:lopez-leon:lavrentovich:tran:23}.
In that case, the director perturbations $\bmu$ would have
dependencies $\bmu=\bmu(x,y,z)$, and the formula for $\div\bmp_0$
would become
\begin{equation*}
  \div \bmp_0 = \Delta\chie E_0
  \bigl[ w_{,x}(x,y,z) \cos q_0 z + w_{,y}(x,y,z) \sin q_0 z \bigr]
  \not= 0 .
\end{equation*}
In such circumstances, the magnetic approximation would \emph{not}
apply, and the local-instability threshold in an electric field would
be \emph{elevated} compared to that predicted by the magnetic-field
analogy.  Periodic instabilities in planar cholesteric films are often
discussed in terms of a \emph{hydrodynamic} instability caused by what
is known as the ``Carr-Helfrich mechanism''---see
\cite[Sec.\,VIII.7]{pikin:91}, \cite[Secs.\,3.10.2 and
4.6.3]{chandrasekhar:92}, or \cite[Sec.\,6.3.3]{degennes:prost:93}.
The discussion here just pertains to instabilities of dielectric
origin producing equilibrium periodic orientational configurations.
These are discussed, for example, in
\cite[Sec.\,4.6.2]{chandrasekhar:92} and
\cite[Sec.\,6.2.2.3]{degennes:prost:93}, and in
\cite[Sec.\,IV.2]{pikin:91}, one can indeed find the suggestion of an
\emph{elevated} electric-field threshold for the case of ``strong
dielectric anisotropy''---the elevating factor (based on an
approximate, coarse-grained free energy) is given there to be
$\sqrt{(\epara+\eperp)/(2\eperp)}$.
% Historical references can be found in the texts cited above.

\subsubsection{Smectic-A phases}

A related periodic instability in a layered system is the
Helfrich-Hurault effect in smectic-A liquid crystals
\cite[Sec.\,IV.3]{pikin:91}, \cite[Sec.\,5.3.3]{chandrasekhar:92},
\cite[Sec.\,7.1.6]{degennes:prost:93}, \cite[Sec.\,6.2]{stewart:04}.
This transition involves a thin film of a smectic-A material of
positive magnetic anisotropy in homeotropic alignment with layers
parallel to the plane of the film and an in-plane magnetic field,
which induces undulations of the layers at a critical field strength.
The system was reconsidered in \cite{bevilacqua:napoli:05} with a
\emph{cross-plane} electric field and a material of \emph{negative}
dielectric anisotropy.  The analysis in that paper indicated that the
critical strength of the electric field should be strictly
\emph{greater} than that predicted by the magnetic-field analogy, by a
factor of $\sqrt{\eperp/\epara}$ (see \cite[Eqns.\,(25) and
(30)]{bevilacqua:napoli:05}).  We show that this is consistent with
our theory.

Consider the system as depicted in
Figure\,\ref{fig:cholesteric-smectic-geoms} right.  The ground-state
director field and electric field are given by
\begin{equation*}
  \nhat_0 = \ez , \quad \bmE_0 = E_0 \ez .
\end{equation*}
At a sufficiently strong electric field, the directors will want to
distort from being parallel to $\bmE$ (since $\epsa<0$) causing the
layers to develop a modulation in the $x$-$y$ plane, with a wavelength
chosen by the system.  The smectic layers are modeled as level
surfaces of a \emph{layer function} $f$, and the liquid-crystal
director is the unit normal to the layers,
\begin{equation*}
  f(x,y,z) = \text{const} , \quad \nhat = \frac{\nabla f}{|\nabla f|} ,
\end{equation*}
with the ground-state layers parallel to the $x$-$y$ plane,
\begin{equation*}
  f_0 = z ~ \Rightarrow ~
  \nhat_0 = \frac{\nabla f_0}{|\nabla f_0|} = \ez .
\end{equation*}
If we choose the $x$ direction for the modulation and assume that the
system remains uniform in the $y$ direction, then the onset of
undulations of the layers will take the form
\begin{equation*}
  f_{\epsilon} = f_0 + \epsilon g(x,z) = z + \epsilon g(x,z) ,
\end{equation*}
where $g$ is periodic in $x$ and vanishes at $z=0$ and $z=d$.  It
follows that
\begin{equation*}
  \nabla f_{\epsilon} = \ez + \epsilon ( g_{,x} \ex + g_{,z} \ez ) =
  \nhat_0 + \epsilon \bmv
\end{equation*}
and
\begin{equation*}
  \nhat_{\epsilon} = \frac{\nabla f_{\epsilon}}{|\nabla f_{\epsilon}|} =
  \frac{\nhat_0+\epsilon\bmv}{|\nhat_0+\epsilon\bmv|} =
  \nhat_0 + \epsilon \bmu + O(\epsilon^2) , \quad
  \bmu = \bfP(\nhat_0) \bmv = g_{,x} \ex .
\end{equation*}
Using these expressions for $\nhat_0$, $\bmE_0$, and $\bmu$, we obtain
% \begin{equation*}
%   \bmp_0 = \Delta\chie \bigl[
%   (\bmu\cdot\bmE_0) \nhat_0 + (\nhat_0\cdot\bmE_0) \bmu \bigr] =
%   \Delta\chie E_0 g_{,x}(x,z) \ex ,
% \end{equation*}
% giving
% \begin{equation*}
%   \div \bmp_0 = \Delta\chie E_0 g_{,xx}(x,z) \not= 0 .
% \end{equation*}
\begin{equation*}
  \bmp_0 = \Delta\chie E_0 g_{,x}(x,z) \ex ~ \Rightarrow ~
  \div \bmp_0 = \Delta\chie E_0 g_{,xx}(x,z) \not= 0 .
\end{equation*}
We see that the \emph{curvature} of the undulating layers leads to a
$\bmp_0$ that is \emph{not} divergence free, and our test predicts an
elevated instability threshold to this modulated phase, consistent
with the analysis in \cite{bevilacqua:napoli:05}.  We note that layer
curvature enters the analysis of the Carr-Helfrich mechanism in a
similar way \cite[Sec.\,4.6.3, Eqn.\,(4.6.32)]{chandrasekhar:92},
\cite[Sec.\,6.3.3, Eqns.\,(6.83) and (6.84)]{degennes:prost:93}.

A similar conclusion is reached for the case of an \emph{in-plane}
electric field with a material of \emph{positive} dielectric
anisotropy (the electric-field version of the original
Helfrich-Hurault instability).  The only difference in this case,
compared to the case discussed above, is in the orientation of the
ground-state electric field.  The ground-state director field,
ground-state electric field, and director perturbations (associated
with the onset of the layer undulations) are now given by
\begin{equation*}
  \nhat_0 = \ez , \quad \bmE_0 = E_0 \ex , \quad \bmu = g_{,x}(x,z) \ex ,
\end{equation*}
from which we obtain
\begin{equation*}
  \bmp_0 = \Delta\chie E_0 g_{,x}(x,z) \ez ~ \Rightarrow ~
  \div \bmp_0 = \Delta\chie E_0 g_{,xz}(x,z) \not= 0 .
\end{equation*}
We conclude that for this system as well, one should observe an
elevation of the instability threshold compared to that predicted by
the magnetic-field analogy.

\subsubsection{Nematics with periodic instabilities}

\label{sec:nematics-with-periodic-instabilities}

Nematic films are also known to develop in-plane orientational
modulations under the influence of magnetic or electric fields in
certain circumstances.  Two examples of this are the periodic
instability studied by Lonberg and Meyer in \cite{lonberg:meyer:85}
and the ``stripe phase'' of Allender, Hornreich, and Johnson
\cite{allender:hornreich:johnson:87}.  Both of these involved magnetic
fields and classical \Freed\ geometries: the splay geometry in the
former case, the bend geometry in the latter.  The experiment studied
in \cite{lonberg:meyer:85} used a polymeric material (characterized by
elastic constants with a small $K_2$ compared to $K_1$) and found that
the ground-state solution became unstable at a magnetic field strength
below that of the classical splay-\Freed\ transition to a
configuration that was periodic in the direction perpendicular to the
plane of the ground-state director field and the applied magnetic
field.  The ``stripe phase'' in \cite{allender:hornreich:johnson:87},
on the other hand, emerged via a secondary bifurcation off the branch
of classical solutions of the bend-\Freed\ transition (at a higher
magnetic field strength) at a temperature near the
nematic-to-smectic-A transition (characterized by $K_1$ small compared
to $K_3$).  One can question if the magnetic-field analogy would be
valid in either of these settings, were the experiments to be done
with \emph{electric} fields instead of magnetic fields.  We show that
in the former case, the answer is ``yes,'' while in the latter case,
it is ``no.''

The periodic instability of \cite{lonberg:meyer:85} is the more
straightforward of the two.  For this we have the splay geometry
(Figure\,\ref{fig:chia-pos-geoms} left) with an equilibrium
configuration that is presumed to remain uniform in $x$ but develop a
periodic modulation in $y$ (in addition to the expected nonuniformity
in the direction of the applied field, the $z$ direction).  The
ground-state director field $\nhat_0$, ground-state electric field
$\bmE_0$, and director perturbations $\bmu$ (which must satisfy
$\nhat_0\cdot\bmu=0$) are given by
\begin{equation*}
  \nhat_0 = \ex , \quad \bmE_0 = E_0 \ez , \quad
  \bmu = v(y,z) \ey + w(y,z) \ez ,
\end{equation*}
with $v$ and $w$ vanishing at $z=0$ and $z=d$.  Using these in
\eqref{eqn:pzero-Ezero}, we obtain
\begin{equation*}
  \bmp_0 = \Delta\chie E_0 w(y,z) \ex ~ \Rightarrow ~
  \div \bmp_0 = 0 .
\end{equation*}
Thus $\bmp_0$ is \emph{divergence free} for all admissible director
perturbations, and we conclude that the magnetic approximation should
be valid: if the experiment were to be done with an electric field
instead of a magnetic field, then the threshold of the instability to
the modulated phase should be correctly given by the formula from the
magnetic-field analogy.

The analysis of the stripe phase is a little more complicated because
of the fact that it enters as a secondary bifurcation.  Thus both the
ground-state director field and the ground-state electric field have
already been distorted by the bend-\Freed\ transition.  In the bend
geometry of Figure\,\ref{fig:chia-pos-geoms} right, they would have the
general forms
\begin{equation*}
  \nhat_0 = n_x^0(z) \ex + n_z^0(z) \ez , \quad
  \bmE_0 = E_0 \ex + E_z(z) \ez ,
\end{equation*}
as seen in Sections~\ref{sec:in-plane-electric-field} and
\ref{sec:analysis-of-E-field-bend}.  The branch that bifurcates off of
this is assumed to contain equilibrium solutions that remain uniform
in $x$ but develop periodicity in $y$.  The perturbation onto this
branch would thus have the general form
\begin{equation*}
  \bmu = u(y,z) \ex + v(y,z) \ey + w(y,z) \ez
\end{equation*}
and be subject to the constraints $\nhat_0\cdot\bmu=0$ and
$\bmu(\cdot,0)=\bmu(\cdot,d)=\bmzero$.  In order to conclude that the
threshold for an electric-field-induced local instability should be
elevated, it is sufficient to find a single perturbation $\bmu$ such
that the field $\bmp_0$ derived from $\nhat_0$, $\bmE_0$, and $\bmu$
is \emph{not} divergence free.  For this, it is enough to consider
the special case $\bmu=v(y,z)\ey$, which gives
\begin{equation*}
  \bmp_0 = \Delta\chie \bigl[ n_x^0(z) E_0 + n_z^0(z) E_z(z) \bigr]
  v(y,z) \ey
\end{equation*}
and
\begin{equation*}
  \div\bmp_0 = \Delta\chie \bigl[ n_x^0(z) E_0 + n_z^0(z) E_z(z) \bigr]
  \frac{\upartial}{\upartial y} v(y,z) \not= 0 .
\end{equation*}
Thus, in contrast to the periodic instability of
\cite{lonberg:meyer:85}, we conclude that in the case of an electric
field, the stripe-phase instability should exhibit an \emph{elevated
  threshold} compared to that predicted by the magnetic-field analogy.

It is difficult to verify these predictions concerning
\cite{lonberg:meyer:85} and \cite{allender:hornreich:johnson:87}, as
there are no explicit formulas for the instability thresholds in
either the magnetic-field or the electric-field cases for either
instability.  In the former case, the ground state is uniform (and
known), and the instability enters with a finite period, but that
period is not known a-priori---see \cite[Sec.\,3.3.2 and Supplementary
Materials]{gartland:21}.  In the latter case, the ground state is
nonuniform (and unknown), and the instability enters with an infinite
period.

\section{Comparison with experiment}

\label{sec:comparison-with-experiment}

The first experiments demonstrating a first-order transition with a
static electric field in the bend geometry were reported in
\cite{frisken:palffy:89}.  While the focus of that work was on the
first-order nature of the transition, one can find evidence of an
elevated local-instability threshold in the data provided there.  The
experimental setup consisted of a cell with two glass substrates
separated by .5\,mm and sandwiched between two stainless steel
electrodes (30.0\,mm x 12.5\,mm x 0.5\,mm) separated by 3.3\,mm.  This
setup was chosen (in lieu of other alternatives) because it provided a
fairly uniform electric field in the sample.

Several different experiments and phenomena are reported in
\cite{frisken:palffy:89}.  Many of the experiments involved a
combination of both an electric field and a magnetic field
(differently oriented), and in addition to regular \Freed\
transitions, instabilities to modulated phases were also observed.
The data that is of use to us here is found in
\cite[Table\,II]{frisken:palffy:89} and corresponds to a classical
electric-field \Freed\ transition in the bend geometry with no
additional magnetic field.  The material used was 5CB at a temperature
of $33.4^{\circ}$C.  Based upon capacitance measurements, the
following hysteresis gap and first-order transition threshold
(taken to be given by $\Vth=(\Vmin+\Vmax)/2$) were reported:
\begin{equation}\label{eqn:Vmin-Vth-Vmax}
  \Vmin = 4.85\,\text{V} , \quad \Vth = 5.1\,\text{V} , \quad
  \Vmax = 5.35\,\text{V} .
\end{equation}
The voltage $\Vmax$ corresponds to the threshold at which the ground
state becomes \emph{locally unstable} (the main focus of our interest
here).

In order to compare the results in this experiment with the modeling
that we have discussed, we require the values of the parameters $K_1$,
$K_3$, $\epara$, and $\eperp$ for this material at this temperature,
not all of which are given in \cite{frisken:palffy:89}---$\epara$,
$\eperp$, and $(K_3-K_1)/K_3$ are given but \emph{not} $K_1$ and $K_3$
individually.  To remedy this, we have used the values for these four
parameters for 5CB at $T=33.4^{\circ}$C obtained from the fitting
formulas in \cite{bogi:faetti:01}.  These values are given by
\begin{equation*}
  K_1 = 3.47 \times 10^{-12}\,\text{J/m} , \quad
  K_3 = 4.22 \times 10^{-12}\,\text{J/m} , \quad
  \epara = 17.5 , \quad \eperp = 7.71 .
\end{equation*}
From these we obtain the following values for the instability
threshold predicted by the magnetic-field analogy and the theoretical
local-instability threshold for the outer asymptotic solution studied
in Section\,\ref{sec:analysis-of-E-field-bend}:
\begin{equation*}%\label{eqn:VH-Vstar}
  V_H = \frac{\pi l}{d} \sqrt{\frac{K_3}{\eps_0\epsa}} =
  4.57\,\text{V} , \quad
  \VBP = \frac{\pi l}{d} \sqrt{\frac{\,\epara}{\,\eperp}}
  \sqrt{\frac{K_3}{\eps_0\epsa}} = 6.89\,\text{V} .
\end{equation*}
We see that the local-instability threshold measured in the experiment
($\Vmax$ in \eqref{eqn:Vmin-Vth-Vmax}) is indeed elevated compared to
$V_H$, though only by roughly 17\% ($(\Vmax-V_H)/V_H\doteq0.171$),
while the theoretical threshold $\VBP$ is elevated by roughly 51\%
($(\VBP-V_H)/V_H\doteq0.508$).  Below we suggest some factors that
could contribute to this discrepancy.

There is no formula from which to calculate the voltage of the
theoretical first-order transition, and the same is true for the width
of the theoretical hysteresis gap.  These must be determined
numerically, which we have done using the same numerical bifurcation
package discussed in Section\,\ref{sec:analysis-of-E-field-bend}.
Those numerical calculations produced the following values for the
limit point $\VLP$ (the lower limit of the hysteresis gap), the
first-order transition threshold $\VUP$ (free-energy crossover), and
the bifurcation point $\VBP$ (the upper limit of the hysteresis gap):
\begin{equation*}%\label{eqn:VLP-VUP-VBP}
  \VLP = 5.99\,\text{V} , \quad
  \VUP = 6.22\,\text{V} , \quad
  \VBP = 6.89\,\text{V} .
\end{equation*}
By comparison, the corresponding values measured in the experiment are
given in \eqref{eqn:Vmin-Vth-Vmax} above.  As already noted, the
ground-state local-instability thresholds $\Vmax$ and $\VBP$ are
elevated from $V_H$ by roughly 17\% and 51\%, while the measured and
theoretical hysteresis gaps differ by
\begin{equation*}
  \Delta V = 0.5\,\text{V} ~~ \text{versus} ~~
  \Delta V^{\text{theor}} = 0.9\,\text{V} .
\end{equation*}

Several factors can be mentioned as potential contributors to the
facts that the instability threshold of the ground state measured in
the experiment is not as elevated as theory predicts and that the
hysteresis gap is not as wide as theory predicts.  The formulas and
numerics that have been developed here (following the original
predictions in \cite{arakelyan:karayan:chilingaryan:84}) are based
upon the assumption that the cell gap is small compared to the width
of the cell, the asymptotic regime $0<d/l\ll1$.  However, the cell
used in the experiments reported in \cite{frisken:palffy:89} did not
have such a small aspect ratio: the actual cell dimensions were
$d=0.5$\,mm and $l=3.3$\,mm, giving $d/l\doteq0.15$.  For such a cell,
it can be expected that the boundary influences are \emph{not}
negligible.  It is also the case that the cell used in the experiments
was quite thick (500\,$\mu$m) in comparison with the typical thickness
of cells used for experiments with liquid crystals (10--50\,$\mu$m),
which would lead to larger fluctuations and poorer alignment.  Thus a
weak-anchoring potential would probably be more realistic than the
infinitely strong homeotropic anchoring assumed in our modeling, and
this would be expected to lead to some reduction of the instability
threshold of the ground state (from that of the strong-anchoring
prediction).  The large fluctuations would also tend to shrink the
measured width of the hysteresis gap (the co-existence region of the
ground state and the distorted director configuration), since
% whether one were to follow the ground state increasing the voltage
% towards the bifurcation point (where the ground state becomes locally
% unstable) or to follow the distorted configuration decreasing the
% voltage towards the limit point (beyond which this solution ceases to
% exist),
fluctuations would drive the system to the competing solution before
reaching either theoretical limit.  We can say that the theory that we
have discussed and the results of the experiment that we have examined
are in qualitative agreement: theory predicts an elevated instability
threshold for the ground state and a first-order phase transition, and
both are observed in the experiment.
% In the experiment, the elevation of the instability threshold is not
% as great as theory predicts (17\% versus 51\%), and the width of the
% hysteresis gap is not as large as theory predicts either (0.5\,V
% versus 0.9\,V).  As discussed above, there are some understandable
% factors contributing to these discrepancies.

\section{Conclusions}

\label{sec:conclusions}

We have studied orientational transitions in nematic liquid crystals
induced by magnetic fields and electric fields, focusing on
differences between the two.  These differences stem from the fact
that magnetic fields can be treated as \emph{uniform} external fields,
unaffected by a liquid-crystal medium, whereas an inhomogeneous
director field will cause \emph{nonuniformity} of an electric field,
in general.  Because of this, magnetic-field-induced transitions are
easier to analyze.  The most basic instabilities of this type are the
classical \Freed\ transitions, and the widely held view is that the
formula for the instability threshold for an electric-field \Freed\
transition can be obtained from that for the magnetic-field transition
in the same geometry by simply replacing the magnetic parameters by
the corresponding electric parameters (the so-called ``magnetic-field
analogy'' or ``magnetic approximation'').  While this is the case for
the splay and twist transitions (with $\epsa>0$), it was shown in
\cite{arakelyan:karayan:chilingaryan:84} that in the case of the
electric-field bend-\Freed\ transition, the local-instability
threshold should be \emph{elevated} from that predicted by the
magnetic-field analogy.

In \cite{gartland:21} we studied the general problem of
electric-field-induced instabilities in nematic systems, confirmed the
results of \cite{arakelyan:karayan:chilingaryan:84}, and derived
necessary conditions for the local stability of general systems
involving coupled director fields and electric fields.  These results
led to the development of a simple test for when the coupling between
a director field and an electric field could lead to such an effect
(altering a local-instability threshold), and they showed that the
effect could only be to \emph{elevate}, never lower, such a threshold.
These results have been reviewed here, where it has been shown how the
classical electric-field \Freed\ transitions fit into the context of
this more general theory.

It is natural to assume the validity of the magnetic approximation in
the setting of classical \Freed\ transitions.  In all of those
geometries, the ground-state director field is \emph{uniform}; so up
to the point of the instability, the electric field is uniform as
well---the magnetic-field analogy is equivalent to modeling the
electric field as always being uniform in the medium.  In general, the
electric field will develop nonuniformity \emph{post transition}, but
the natural assumption would be that this would not affect the
critical voltage at which the instability occurs.  The mechanism that
underlies the threshold-elevating effect is related to the mutual
influence of the director field $\nhat$ and the electric field $\bmE$
and to changes in dielectric polarization that take place at the onset
of an electric-field-induced instability ($\bmP-\bmP_0$ as a function
of $\delta\nhat$).  The reorientation of the director field at a
configurational transition will cause changes in the induced
polarization and, in general, changes in the local electric field.  In
certain geometries, these changes have an effect on $\div\bmP$ that is
\emph{first order} in $\delta\nhat$, while in other geometries, the
effect is \emph{higher order}.  First-order effects elevate
instability thresholds, while higher order effects do not---they only
produce quantitative differences \emph{post transition}.  We have
explored these ideas in detail.

The issues are similar in spirit to flexoelectricity (polarization
effects related to director distortion), but no flexoelectric
coefficients are involved.  The effect is purely one of induced
polarization in a linear dielectric.  The inclusion of flexoelectric
terms in the free-energy density affects some of the analysis and
leads to additional contributions to the total polarization $\bmP$ and
to the right-hand side of \eqref{eqn:div-grad-psi}, but these do not
alter the local-instability thresholds of the classical electric-field
\Freed\ transitions, for example (see \cite[Sec.\,5.1]{gartland:21}).
The way in which the elevation of an instability threshold is effected
can be understood from the point of view of \emph{energetics} or from
that of \emph{torque balance}.  From the \emph{energy} point of view,
the changes in the electric field that take place at the onset of such
an instability increase the magnitude of the field and add to the
electrostatic component of the free energy, which increases the
barrier that must be overcome to destabilize the ground-state director
configuration.  While from the \emph{force} point of view, the
addition to the electric field caused by these changes is in the
direction of the ground-state director field and therefore is
\emph{aligning} and works against the dielectric torque that is trying
to rotate the ground-state director field.

Whether an electric-field \Freed\ transition has an elevated threshold
or not is not a simple matter of in-plane versus cross-plane electric
field: both the bend and twist transitions have in-plane fields, yet
the bend transition has an elevated threshold, while the twist
transition does not.  Nor is it a simple matter of whether or not
nonuniformity develops in the electric field post transition: both the
splay and bend transitions have nonuniform electric fields past the
instability threshold, yet the bend transition has an elevated
threshold, while the splay transition does not.  The issue hinges on
\emph{electrostatic equilibrium} ($\div\bmD=0$,
$\bmD=\epstensor(\nhat)\bmE$).  If at the onset of an instability,
$\div\bmD=0$ is maintained to first order in $\delta\nhat$ with an
\emph{unperturbed} ground-state electric field $\bmE=\bmE_0$, then
there will \emph{not} be an elevated threshold.  If, on the other
hand, a perturbation of the ground-state electric field is required to
preserve $\div\bmD=0$ to first order in $\delta\nhat$, then an
elevation of the local instability threshold \emph{will} occur.  The
phenomenon occurs in more general systems than just classical \Freed\
transitions.

A simple test to determine whether a transition will have an elevated
local-instability threshold or not was given in
Section\,\ref{sec:application-to-classical-Freed}.  This test depends
on the ground-state director field, the ground-state electric field,
and the admissible director perturbations (all of which depend on the
geometry and symmetry assumptions of the particular system).  This
test was applied to several electric-field-induced instabilities,
including both classical and generalized \Freed\ transitions.  A few
of the identified electric-field transitions that manifest this
``anomalous behavior'' have already been analyzed, including the
bend-\Freed\ transition with $\epsa>0$
\cite{arakelyan:karayan:chilingaryan:84}, the splay-\Freed\ transition
with $\epsa<0$ \cite{arakelyan:karayan:chilingaryan:84}, and the
electric-field Helfrich-Hurault instability in smectic-A films with
$\epsa<0$ \cite{bevilacqua:napoli:05}.  Other systems predicted by our
theory and test to have elevated thresholds, but which have not yet
been analyzed (to the best of our knowledge), include periodic
instabilities in planar cholesteric films and the modulated ``stripe
phase'' in nematics.  The test correctly predicts that there should be
\emph{no elevation} for the thresholds of the other classical \Freed\
transitions (splay transition with $\epsa>0$, bend transition with
$\epsa<0$, and twist transition with $\epsa>0$ or $\epsa<0$), as well
as for twisted nematics with a small number of twists (e.g., TNC or
STN switching thresholds).  It also predicts that the
local-instability threshold should \emph{not} be elevated for the
periodic instability studied in \cite{lonberg:meyer:85}, though this
has not been analyzed either (as far as we know).  We note that the
test does not rely on the ground-state director field and electric
field being uniform, as evidenced by the applications in
Sections~\ref{sec:twisted-nematics-and-cholesterics} and
\ref{sec:nematics-with-periodic-instabilities}.

In summary, in certain systems the nature of the coupling between the
equilibrium liquid-crystal director field and the electric field can
cause an elevation of the threshold of local instability compared to
that predicted by the magnetic-field analogy.  We have explored this
carefully and have explained why it happens and how to identify when
it can happen.  The effect and analysis extend beyond classical
\Freed\ transitions in nematics to other configurations (e.g., twisted
nematics, secondary bifurcations in nematics), to other liquid crystal
phases (e.g., cholesteric, smectic~A), as well as to periodic
instabilities.  The analysis here was done using the macroscopic
continuum theory of Oseen, Zocher, and Frank, but the development
should be similar and the conclusions the same in the mesoscopic
Landau-de\,Gennes continuum theory.

\section*{Acknowledgments}

The author is grateful to O.~D.~Lavrentovich and P.~Palffy-Muhoray for
helpful discussions.  Part of this work was performed at the Isaac
Newton Institute for Mathematical Sciences (University of Cambridge),
and the author thanks that organization for its support.

\section*{Disclosure statement}

No potential conflict of interest was reported by the author.

\section*{Funding}

This work was supported in part by the Division of Mathematical
Sciences, U.S.~National Science Foundation [grant number
DMS-1211597].

\section*{ORCID}

\textit{Eugene C. Gartland, Jr.}, \url{https://orcid.org/0000-0002-6956-0538}

%%%%%%%%%% REFERENCES %%%%%%%%%%

%\nocite{*}           % list contents of paper.bib in References (for checking)

\bibliography{paper.bib}

\begin{thebibliography}{31}
\providecommand{\natexlab}[1]{#1}
\providecommand{\url}[1]{\texttt{#1}}
\expandafter\ifx\csname urlstyle\endcsname\relax
  \providecommand{\doi}[1]{doi: #1}\else
  \providecommand{\doi}{doi: \begingroup \urlstyle{rm}\Url}\fi

\bibitem[Pikin(1991)]{pikin:91}
S.~A. Pikin.
\newblock \emph{Structural Transformations in Liquid Crystals}.
\newblock Gordon and Breach Science Publishers, New York, 1991.
\newblock Translated from the Russian by Michael E. Alferieff.

\bibitem[Chandrasekhar(1992)]{chandrasekhar:92}
S.~Chandrasekhar.
\newblock \emph{Liquid Crystals}.
\newblock Cambridge University Press, Cambridge, 2nd edition, 1992.

\bibitem[de~Gennes and Prost(1993)]{degennes:prost:93}
Pierre~Gilles de~Gennes and Jacques Prost.
\newblock \emph{The Physics of Liquid Crystals}.
\newblock Clarendon Press, Oxford, 2nd edition, 1993.

\bibitem[Virga(1994)]{virga:94}
Epifanio~G. Virga.
\newblock \emph{Variational Theories for Liquid Crystals}.
\newblock Chapman \& Hall, London, 1994.

\bibitem[Stewart(2004)]{stewart:04}
Iain~W. Stewart.
\newblock \emph{The Static and Dynamic Continuum Theory of Liquid Crystals}.
\newblock Taylor \& Francis, London, 2004.

\bibitem[Arakelyan et~al.(1984)Arakelyan, Karayan, and
  Chilingaryan]{arakelyan:karayan:chilingaryan:84}
S.~M. Arakelyan, A.~S. Karayan, and Yu.~S. Chilingaryan.
\newblock Fr\'{e}edericksz transition in nematic liquid crystals in static and
  light fields: general features and anomalies.
\newblock \emph{Sov. Phys. Dokl.}, 29\penalty0 (3):\penalty0 202--204, 1984.
\newblock Translation of Dokl.\ Akad.\ Nauk SSSR 275 (1), 52--55 (March 1984).

\bibitem[Gartland(2021)]{gartland:21}
Eugene~C. Gartland, Jr.
\newblock Electric-field-induced instabilities in nematic liquid crystals.
\newblock \emph{SIAM J. Appl. Math.}, 81\penalty0 (2):\penalty0 304--334, 2021.
\newblock doi: \href{https://doi.org/10.1137/20M134349X}{10.1137/20M134349X}.

\bibitem[Gruler and Meier(1972)]{gruler:meier:72}
Hans Gruler and Gerhard Meier.
\newblock Electric field-induced deformations in oriented liquid crystals of
  the nematic type.
\newblock \emph{Mol. Cryst. Liq. Cryst.}, 16\penalty0 (4):\penalty0 299--310,
  1972.
\newblock doi: \href{https://doi.org/10.1080/15421407208082793}{10.1080/15421407208082793}.

\bibitem[Gruler et~al.(1972)Gruler, Scheffer, and
  Meier]{gruler:scheffer:meier:72}
Hans Gruler, Terry~J. Scheffer, and Gerhard Meier.
\newblock Elastic constants of nematic liquid crystals: I. {T}heory of the
  normal deformation.
\newblock \emph{Z. Naturforsch. A}, 27\penalty0 (6):\penalty0 966--976, 1972.
\newblock doi: \href{https://doi.org/10.1515/zna-1972-0613}{10.1515/zna-1972-0613}.

\bibitem[Deuling(1972)]{deuling:72}
Heinz~J. Deuling.
\newblock Deformation of nematic liquid crystals in an electric field.
\newblock \emph{Mol. Cryst. Liq. Cryst.}, 19\penalty0 (2):\penalty0 123--131,
  1972.
\newblock doi: \href{https://doi.org/10.1080/15421407208083858}{10.1080/15421407208083858}.

\bibitem[Frisken and Palffy-Muhoray(1989{\natexlab{a}})]{frisken:palffy:89}
B.~J. Frisken and P.~Palffy-Muhoray.
\newblock Electric-field-induced twist and bend {F}reedericksz transitions in
  nematic liquid crystals.
\newblock \emph{Phys. Rev. A}, 39\penalty0 (3):\penalty0 1513--1518,
  1989{\natexlab{a}}.
\newblock doi: \href{https://doi.org/10.1103/PhysRevA.39.1513}{10.1103/PhysRevA.39.1513}.

\bibitem[Frisken and Palffy-Muhoray(1989{\natexlab{b}})]{frisken:palffy:89b}
B.~J. Frisken and P.~Palffy-Muhoray.
\newblock Effects of a transverse electric field in nematics: {I}nduced
  biaxiality and the bend {F}r\'{e}edericksz transition.
\newblock \emph{Liquid Crystals}, 5\penalty0 (2):\penalty0 623--631,
  1989{\natexlab{b}}.
\newblock doi: \href{https://doi.org/10.1080/02678298908045413}{10.1080/02678298908045413}.

\bibitem[Richards(2006)]{richards:06}
Gregory~P. Richards.
\newblock Numerical modeling and stability analysis of the electric-field
  bend-{F}r\'eedericksz transition for nematic liquid crystal cells.
\newblock {M}asters {T}hesis, {A}pplied {M}athematics, Kent State University,
  Kent, OH, USA, August 2006.

\bibitem[Kini(1990)]{kini:90}
U.~D. Kini.
\newblock Discontinuous orientational changes in nematics: {E}ffects of electric
  and magnetic fields.
\newblock \emph{Liquid Crystals}, 8\penalty0 (6):\penalty0 745--763, 1990.
\newblock doi: \href{https://doi.org/10.1080/02678299008047386}{10.1080/02678299008047386}.

\bibitem[Kinderlehrer and Ou(1992)]{kinderlehrer:ou:92}
David Kinderlehrer and Biao Ou.
\newblock Second variation of liquid crystal energy at $x/|x|$.
\newblock \emph{Proc. Royal Soc. Lond. A}, 437\penalty0 (1900):\penalty0
  475--487, 1992.
\newblock doi: \href{https://doi.org/10.1098/rspa.1992.0074}{10.1098/rspa.1992.0074}.

\bibitem[Rosso et~al.(2004)Rosso, Virga, and Kralj]{rosso:virga:kralj:04}
Riccardo Rosso, Epifanio~G. Virga, and Samo Kralj.
\newblock Local elastic stability for nematic liquid crystals.
\newblock \emph{Phys. Rev. E}, 70\penalty0 (1):\penalty0 011710, 2004.
\newblock doi: \href{https://doi.org/10.1103/PhysRevE.70.011710}{10.1103/PhysRevE.70.011710}.

\bibitem[Frisken and Palffy-Muhoray(1989{\natexlab{c}})]{frisken:palffy:89c}
B.~J. Frisken and P.~Palffy-Muhoray.
\newblock Freedericksz transitions in nematic liquid crystals: The effects of
  an in-plane electric field.
\newblock \emph{Phys. Rev. A}, 40\penalty0 (10):\penalty0 6099--6102,
  1989{\natexlab{c}}.
\newblock doi: \href{https://doi.org/10.1103/PhysRevA.40.6099}{10.1103/PhysRevA.40.6099}.

\bibitem[Schiller(1990)]{schiller:90}
P.~Schiller.
\newblock Equilibrium structures of planar nematic and cholesteric films in
  electric fields.
\newblock \emph{Phase Transitions}, 29\penalty0 (2):\penalty0 59--83, 1990.
\newblock doi: \href{https://doi.org/10.1080/01411599008207944}{10.1080/01411599008207944}.

\bibitem[Self et~al.(2002)Self, Please, and Sluckin]{self:please:sluckin:02}
R.~H. Self, C.~P. Please, and T.~J. Sluckin.
\newblock Deformation of nematic liquid crystals in an electric field.
\newblock \emph{Eur. J. Appl. Math.}, 13:\penalty0 1--23, 2002.
\newblock doi: \href{https://doi.org/10.1017/S0956792501004740}{10.1017/S0956792501004740}.

\bibitem[Napoli(2006)]{napoli:06}
Gaetano Napoli.
\newblock Weak anchoring effects in electrically driven {F}reedericksz
  transitions.
\newblock \emph{J. Phys. A: Math. Gen.}, 39\penalty0 (1):\penalty0 11--31,
  2006.
\newblock doi: \href{https://doi.org/10.1088/0305-4470/39/1/002}{10.1088/0305-4470/39/1/002}.

\bibitem[Nayfeh(1981)]{nayfeh:81}
Ali~Hasan Nayfeh.
\newblock \emph{Introduction to Perturbation Techniques}.
\newblock John Wiley \& Sons, New York, 1981.

\bibitem[Leslie(1970)]{leslie:70}
F.~M. Leslie.
\newblock Distortion of twisted orientation patterns in liquid crystals by
  magnetic fields.
\newblock \emph{Mol. Cryst. Liq. Cryst.}, 12\penalty0 (1):\penalty0 57--72,
  1970.
\newblock doi: \href{https://doi.org/10.1080/15421407008082760}{10.1080/15421407008082760}.

\bibitem[Schadt and Helfrich(1971)]{schadt:helfrich:71}
M.~Schadt and W.~Helfrich.
\newblock Voltage-dependent optical activity of a twisted nematic liquid
  crystal.
\newblock \emph{Appl. Phys. Lett.}, 18\penalty0 (4):\penalty0 127--128, 1971.
\newblock doi: \href{https://doi.org/10.1063/1.1653593}{10.1063/1.1653593}.

\bibitem[Fraser(1978)]{fraser:78}
C.~Fraser.
\newblock Theoretical investigation of {F}r\'eedericksz transitions in twisted
  nematics with surface tilt.
\newblock \emph{J. Phys. A: Math. Gen.}, 11\penalty0 (7):\penalty0 1439--1448,
  1978.
\newblock doi: \href{https://doi.org/10.1088/0305-4470/11/7/030}{10.1088/0305-4470/11/7/030}.

\bibitem[Raynes(1986)]{raynes:86}
E.~P. Raynes.
\newblock The theory of supertwist transitions.
\newblock \emph{Mol. Cryst. Liq. Cryst.}, 4\penalty0 (1):\penalty0 1--8, 1986.
\newblock URL \url{https://www.tandfonline.com/doi/abs/10.1080/01406566.1986.10766872}.

\bibitem[Schiller(1989)]{schiller:89}
P.~Schiller.
\newblock Perturbation theory for planar nematic twisted layers.
\newblock \emph{Liq. Cryst.}, 4\penalty0 (1):\penalty0 69--78, 1989.
\newblock doi: \href{https://doi.org/10.1080/02678298908028959}{10.1080/02678298908028959}.

\bibitem[Blanc et~al.(2023)Blanc, Durey, Kamien, Lopez-Leon, Lavrentovich, and
  Tran]{blanc:durey:kamien:lopez-leon:lavrentovich:tran:23}
Christophe Blanc, Guillaume Durey, Randall~D. Kamien, Teresa Lopez-Leon,
  Maxim~O. Lavrentovich, and Lisa Tran.
\newblock Helfrich-{H}urault elastic instabilities driven by geometrical
  frustration.
\newblock \emph{Rev. Mod. Phys.}, 95\penalty0 (1):\penalty0 015004, 2023.
\newblock doi: \href{https://doi.org/10.1103/RevModPhys.95.015004}{10.1103/RevModPhys.95.015004}.

\bibitem[Bevilacqua and Napoli(2005)]{bevilacqua:napoli:05}
G.~Bevilacqua and G.~Napoli.
\newblock Re-examination of the {H}elfrich-{H}urault effect in smectic-{$A$}
  liquid crystals.
\newblock \emph{Phys. Rev. E}, 72\penalty0 (4):\penalty0 041708, 2005.
\newblock doi: \href{https://doi.org/10.1103/PhysRevE.72.041708}{10.1103/PhysRevE.72.041708}.

\bibitem[Lonberg and Meyer(1985)]{lonberg:meyer:85}
Franklin Lonberg and Robert~B. Meyer.
\newblock New ground state for the splay-{F}r\'{e}edericksz transition in a
  polymer nematic liquid crystal.
\newblock \emph{Phys. Rev. Lett.}, 55\penalty0 (7):\penalty0 718--721, 1985.
\newblock doi: \href{https://doi.org/10.1103/PhysRevLett.55.718}{10.1103/PhysRevLett.55.718}.

\bibitem[Allender et~al.(1987)Allender, Hornreich, and
  Johnson]{allender:hornreich:johnson:87}
D.~W. Allender, R.~M. Hornreich, and D.~L. Johnson.
\newblock Theory of the stripe phase in bend-{F}r\'{e}edericksz-geometry
  nematic films.
\newblock \emph{Phys. Rev. Lett.}, 59\penalty0 (23):\penalty0 2654--2657, 1987.
\newblock doi: \href{https://doi.org/10.1103/PhysRevLett.59.2654}{10.1103/PhysRevLett.59.2654}.

\bibitem[Bogi and Faetti(2001)]{bogi:faetti:01}
A.~Bogi and S.~Faetti.
\newblock Elastic, dielectric and optical constants of
  $4'$-pentyl-4-cyanobiphenyl.
\newblock \emph{Liquid Crystals}, 28\penalty0 (5):\penalty0 729--739, 2001.
\newblock doi: \href{https://doi.org/10.1080/02678290010021589}{10.1080/02678290010021589}.

\end{thebibliography}

%%%%%%%%%% APPENDICES %%%%%%%%%%

%\appendix

% \section{Perturbation analysis of the electric-field splay-\Freed\ transition}

% \label{app:pert-anal-of-E-field-splay}

% ???

% \section{Asymptotic analysis for the bend geometry}

% ???

% \subsection{Full coupled system}

% ???

% \subsection{Scaling analysis}

% ???

% \subsection{Outer problem and solution for the electric potential}

% ???

\end{document}